\documentclass[aps,prd,onecolumn,groupedaddress,showpacs,nofootinbib,amssymb]{revtex4}
\usepackage[dvips]{graphicx}
\usepackage{amssymb}
\usepackage{amsmath}
\usepackage{graphicx,color}
\usepackage{amsfonts}
\usepackage{bm}
\usepackage{cancel}
\usepackage{comment}

\newcommand\be{\begin{equation}}
\newcommand\ee{\end{equation}}

\allowdisplaybreaks[4]

\begin{document}

\title{Dynamical Dark Energy from $F(R)$ Gravity Models Unifying Inflation with Dark Energy: Confronting the Latest Observational Data}
\author{S.D. Odintsov$^{1,2,3,4}$}\email{odintsov@ice.csic.es}
\author{V.K. Oikonomou,$^{4,5}$}\email{voikonomou@gapps.auth.gr;v.k.oikonomou1979@gmail.com}
\author{G.S. Sharov$^{6,7}$}\email{sharov.gs@tversu.ru}
\affiliation{$^{1)}$ Institute of Space Sciences (ICE, CSIC) C. Can Magrans s/n, 08193 Barcelona, Spain \\
$^{2)}$ ICREA, Passeig Luis Companys, 23, 08010 Barcelona, Spain\\
$^{3)}$Institut d'Estudis Espacials de Catalunya (IEEC), 08860
Castelldefels, Barcelona, Catalonia, Spain\\
$^{4)}$L.N. Gumilyov Eurasian National University - Astana,
010008, Kazakhstan \\
$^{5)}$Department of Physics, Aristotle University of
Thessaloniki, Thessaloniki 54124, Greece\\
$^{6)}$Tver state university, Sadovyj per. 35, 170002 Tver,
Russia\\
$^{7)}$International Laboratory for Theoretical Cosmology, Tomsk
State University of Control Systems and Radioelectronics (TUSUR),
634050 Tomsk, Russia}

\tolerance=5000

\begin{abstract}
A class of viable $F(R)$ gravity models which can provide a
unified description of inflation with the dark energy era is
confronted with the latest observational data on the dark energy
era. These models have the unique characteristic that the de
Sitter scalaron mass in the Einstein frame counterpart theory is a
monotonic function of the curvature, which renders them viable
descriptions for both the inflationary and the late-time
acceleration eras. We also compare these models with other
well-known viable $F(R)$ gravity models and with the
$\Lambda$-Cold-Dark-Matter model. As we show, the most
phenomenologically successful models are those which deviate
significantly from the  $\Lambda$-Cold-Dark-Matter model. Also
some of the models presented, provide a statistically favorable
description of the dark energy eras, compared with the exponential
$F(R)$ gravity model and of course compared with the
$\Lambda$-Cold-Dark-Matter model. All the models we present in
this article are confronted with the observational data from the
Planck collaboration, the Pantheon plus data from Type Ia
supernovae, the two rounds of observations of the Dark Energy
Spectroscopic Instrument, data from baryon acoustic oscillations
and the Hubble constant measurements.  
 As we show,
two of the models are statistically favorable by the data.
\end{abstract}


\maketitle

\section{Introduction}

The $\Lambda$-Cold-Dark-Matter model ($\Lambda$CDM) is the
benchmark of cosmology to date, since it fits perfectly the cosmic
microwave background (CMB) radiation data \cite{Planck2018} and
aligns with a successful structure formation. However, data coming
from the Pantheon plus catalog of Type Ia supernovae (SNe Ia)
\cite{PantheonP:2022} and also baryon acoustic oscillations (BAO)
data coming from the two rounds of observations of the Dark Energy
Spectroscopic Instrument (DESI) \cite{DESI:2024,DESI:2025zgx}
indicate that there are shortcomings in the $\Lambda$CDM
description of the late-time era. To these data, one must add the
Hubble constant measurements by SH0ES group \cite{Riess2021} and
hence the $\Lambda$CDM is currently strongly challenged. The
latest strong indication that the $\Lambda$CDM might not be an
efficient description of the late-time era, came from the second
round of observations of DESI \cite{DESI:2025zgx}, which indicated
that the dark energy is dynamical and more importantly it evolves
from a phantom to a quintessence equation of state (EoS). This
result is confirmed with a statistical confidence up to
$4.2\sigma$, when BAO data are taken into account. The
$\Lambda$CDM is a general relativistic framework, and thus a
dynamical phantom evolution might be difficult to accommodate in
the context of general relativity without resorting to the use of
tachyon scalar fields which evolve to quintessence at late times.
This is a difficult task to theoretically achieve, but in the
literature there are various ways of describing such dynamical
evolutions in order to evade the $\Lambda$CDM problems, see for
example
\cite{OdintsovSGS:2024,Dai:2020rfo,He:2020zns,Nakai:2020oit,DiValentino:2020naf,Agrawal:2019dlm,Yang:2018euj,Ye:2020btb,Vagnozzi:2021tjv,
Desmond:2019ygn,Hogas:2023pjz,OColgain:2018czj,Vagnozzi:2019ezj,
Krishnan:2020obg,Colgain:2019joh,Vagnozzi:2021gjh,Lee:2022cyh,Krishnan:2021dyb,Ye:2021iwa,Ye:2022afu,Verde:2019ivm,Menci:2024rbq,Adil:2023ara,
Reeves:2022aoi,Ferlito:2022mok,Vagnozzi:2021quy,DiValentino:2020evt,DiValentino:2019ffd,Sabogal:2024yha,Giare:2024smz,DiValentino:2025sru,Odintsov:2025kyw,vanderWesthuizen:2025iam,Paliathanasis:2025dcr}.
Modified gravity \cite{reviews1,reviews2,reviews3,reviews4} plays
a prominent role in these $\Lambda$CDM emulator theories, which
keep the good phenomenological features of the $\Lambda$CDM, while
they provide a remedy for the shortcomings of the $\Lambda$CDM
model. In this work we shall focus on the most characteristic
class of modified gravity models, the $F(R)$ gravity theories,
which contain higher order Ricci scalar corrections in the
Lagrangian
\cite{Nojiri:2003ft,Capozziello:2005ku,Hwang:2001pu,Song:2006ej,Faulkner:2006ub,Olmo:2006eh,Sawicki:2007tf,Faraoni:2007yn,Carloni:2007yv,
Nojiri:2007as,Deruelle:2007pt,Appleby:2008tv,Dunsby:2010wg,Oikonomou:2025qub}.
We shall analyze several classes of models which remarkably
provide a unified description of the early and late-time era.
These models generate a scalaron mass in the Einstein frame
counterpart theory that is monotonically increasing with the
curvature, and this feature makes them viable theories for both
inflation and late-time dynamics, see \cite{Oikonomou:2025qub} for
more details on this. In our analysis we include three such models
and we demonstrate that some of these phenomenological models are
fully compatible with the observational data, including the second
round of the DESI data. All these models are $\Lambda$CDM
emulators at late times, but more importantly these models are
compatible with the data and provide a natural framework to
realize phantom to quintessence transitions without resorting to
tachyon fields.

Before getting to the core of our analysis, let us fix the
background spacetime which shall be used in this article, and we
assume that it is that of a flat Friedmann-Robertson-Walker (FRW)
spacetime with line element,
\begin{equation}
\label{JGRG14} ds^2 = - dt^2 + a(t)^2 \sum_{i=1,2,3}
\left(dx^i\right)^2\, ,
\end{equation}
where $a(t)$ is the scale factor and the Hubble rate is
$H=\frac{\dot{a}}{a}$.

The paper is organized as follows: in section II, the dynamical
equations for $F(R)$ gravity late-time description are presented
using two different equivalent forms. In section III three $F(R)$
models are presented, which are $\Lambda$CDM emulators at late
times, while generate a viable inflationary era. In section IV
these three models are confronted with SNe Ia, $H(z)$, CMB and BAO
DESI observational data and compared with the exponential $F(R)$
and $\Lambda$CDM models. Finally, the conclusions follow at the
end of the article.

\section{$F(R)$ Gravity Unifying Inflation with the Dark Energy Era}
\label{Dynamics}

In this section we shall discuss the theoretical framework of the
$F(R)$ gravity models which we shall confront with the latest dark
energy data. These models are theoretically motivated by the fact
that these stem from a theoretical basis which unifies inflation
and the dark energy era with the same model. To start with, let us
discuss the $F(R)$ gravity theoretical framework, the
gravitational action of which in the presence of perfect matter
fluids is,
\begin{equation}
\label{actionde} \centering
\mathcal{S}=\int{d^4x\sqrt{-g}\left(\frac{F(R)}{2\kappa^2}+\mathcal{L}_m\right)}\, ,
\end{equation}
where $\mathcal{L}_m$ stands for the Lagrangian density of the
perfect matter fluids which are present. We shall choose the
$F(R)$ to be in the form,
\begin{equation}\label{FR}
    F(R)=R+f(R) .
\end{equation}
hence by varying the gravitational action (\ref{actionde}) with
respect to the metric, we obtain the field equations,
\begin{equation} \label{Friedman}
    3 F_R H^2=\kappa^2 \rho_m + \frac{F_R R - F}{2} -3H \dot F_R \, ,
\end{equation}
\begin{equation} \label{Raycha}
    -2 F_R \dot H = \kappa^2 (\rho_m + R_m) + \ddot F -H \dot F \, ,
\end{equation}
where $F_R = \frac{\partial F}{\partial R}$ and the ``dot'' in the
equations above denotes the derivative with respect to the cosmic
time. In addition, $\rho_m$ and $P_m$ stand for the energy density
and the pressure of the perfect matter fluids. We can rewrite the
field equations (\ref{Friedman}), (\ref{Raycha}) in the
Einstein-Hilbert gravity form for a flat FRW spacetime in the
following way,
\begin{equation} \label{Friedtot}
    3 H^2= \kappa^2 \rho_\mathrm{tot}  \ ,
\end{equation}
\begin{equation} \label{Raychtot}
   -2 \dot H =\kappa^2 (\rho_\mathrm{tot} + P_\mathrm{tot}) \ ,
\end{equation}
where $\rho_\mathrm{tot}$ stands for the total energy density of
the total cosmological fluid and $P_\mathrm{tot}$ stands for the
corresponding total pressure. The total effective cosmological
fluid consists of three parts, one corresponding to the cold dark
matter ($\rho_m$), one corresponding to the radiation part
($\rho_r$) and finally one corresponding to the geometric part
($\rho_{DE}$), generated by $F(R)$ gravity. Therefore, we have,
$\rho_\mathrm{tot}=\rho_m + \rho_r + \rho_{DE}$ and in addition,
$P_\mathrm{tot}=P_m + P_r + P_{DE}$. The geometric fluid controls
the evolution at early and late-times, and the energy density and
effective pressure of this fluid are,
\begin{equation}\label{rDE}
    \rho_{DE}=\frac{F_R R - F}{2} + 3 H^2 (1-F_R)-3H \dot F_R \, ,
\end{equation}
\begin{equation}\label{PDE}
    P_{DE}=\ddot F -H \dot F +2 \dot H (F_R -1) - \rho_{DE} \, .
\end{equation}
Using the redshift,
\begin{equation}
    1+z=\frac{1}{a}\, ,
\end{equation}
as a dynamical variable for the cosmic evolution, and by
introducing the statefinder function $y_H (z)$
\cite{Hu:2007nk,Bamba:2012qi,reviews1},
\begin{equation} \label{yhdef}
    y_H(z)=\frac{\rho_{DE}}{\rho_m^{(0)}}=\frac{H^2}{m_s^2}-(1+z)^3-X_r (1+z)^4 ,
\end{equation}
we can quantify the dark energy evolution in terms of $y_H (z)$.
Also recall that $\rho_m^{(0)}$ stands for the energy density of
the cold dark matter at present time, and in addition,
$m_s^2=\frac{\kappa^2 \rho_m^{(0)}}{3}=H_0^2 \Omega_m=1.37 \times
10^{-67} eV^2$. Finally, the radiation to matter ratio $X_r$ is
equal to,
   \begin{equation} \label{Xrm}
X_r=\frac{\rho_r^{(0)}}{\rho_m^{(0)}}=2.9656\cdot10^{-4}
\end{equation}
Combining of Eqs. (\ref{Friedtot}) , (\ref{FR}) and (\ref{yhdef}),
the Friedmann equation (\ref{Friedman}) can takes the following
form,
\begin{equation} \label{difyh}
    \frac{d^2 y_H}{dz^2} + J_1 \frac{d y_H}{dz} + J_2 y_H + J_3=0 \, ,
\end{equation}
where the dimensionless functions $J_1$ , $J_2$ , $J_3 $ are
defined as follows,
\begin{equation} \label{J1}
    J_1 = \frac{1}{(z+1)} \Big( -3-\frac{1}{y_H + (z+1)^3 + X_r (z+1)^4}  \frac{1-F_R}{6 m_s^2F_{RR}}\Big) \, ,
\end{equation}
\begin{equation} \label{J2}
    J_2 = \frac{1}{(z+1)^2} \Big( \frac{1}{y_H + (z+1)^3 +X_r (z+1)^4} \frac{2-F_R}{3 m_s^2 F_{RR}}\Big) \, ,
\end{equation}
\begin{equation} \label{J3}
    J_3 = -3(z+1) - \frac{(1-F_R)((z+1)^3 + 2X_r (z+1)^4) + (R-F)/(3 m_s^2)}{(z+1)^2 (y_H + (z+1)^3 +X_r (z+1)^4)} \frac{1}{6 m_s^2 F_{RR}}\, ,
\end{equation}
and also $F_{RR}=\frac{\partial^2 F}{\partial R^2}$. Moreover, the
Ricci scalar is,
\begin{equation}\label{ft10newadd}
R=12H^2-6HH_{z}(1+z)\, ,
\end{equation}
or expressed in terms of the statefinder $y_H$,
\begin{equation}\label{neweqnrefricciyH}
R(z)=3\,m_s^2\left(-(z+1)\,\frac{d y_H(z)}{dz} + 4 y_H(z) + (1+z)^3\right)\, .
\end{equation}
One can study the late-time dynamics by solving the differential
equation Eq. (\ref{difyh}) numerically, for appropriate redshift
intervals. The initial conditions at the redshift $z_f=10$ are
\cite{Bamba:2012qi},
\begin{equation}\label{initialcond}
    y_H (z_f) = \frac{ \Lambda}{3 m_s^2} \Big( 1 + \frac{1+z_f}{1000} \Big)\;, \qquad \frac{d y_H(z)}{dz} \Big |_{z=z_f} = \frac{1}{1000} \frac{ \Lambda}{3 m_s^2},
\end{equation}
where $\Lambda \simeq 11.895 \times 10^{-67}$ eV${}^2$ is the
cosmological constant. We can express the physical quantities
relevant for cosmology in terms of the statefinder function
$y_H(z)$ as follows,
\begin{equation}\label{hubblefr}
H(z)=m_s\sqrt{y_H(z)+(1+z)^{3}+X_r (1+z)^4}\, .
\end{equation}
and also the Ricci scalar is, Eq.~(\ref{neweqnrefricciyH})
 \begin{equation}\label{curvature}
    R(z)=3 m_s^2 \Big( 4 y_H(z) -(z+1) \frac{d y_H (z)}{dz} + (z+1)^3 \Big),
\end{equation}
and also the dark energy density parameter $\Omega_{DE}(z)$ is
defined as,
\begin{equation}\label{OmegaDE}
    \Omega_{DE}(z)=\frac{y_H(z)}{y_H(z)+(z+1)^3 + X_r (z+1)^4}\, ,
\end{equation}
and the dark energy EoS parameter is expressed as follows,
\begin{equation}\label{EoSDE}
    \omega_{DE}(z)=-1+\frac{1}{3}(z+1)\frac{1}{y_H(z)}\frac{d y_H(z)}{dz},
\end{equation}
while the total EoS parameter is equal to,
\begin{equation}\label{EoStot}
    \omega_\mathrm{tot}(z)=\frac{2(z+1)H'(z)}{3H(z)}-1 \, .
\end{equation}
Furthermore, the deceleration parameter is equal to,
\begin{equation}\label{declpar}
    q(z)=-1-\frac{\dot H}{H^2}=-1-(z+1)\frac{H'(z)}{H(z)},
\end{equation}
with the ``prime'' this time, denoting differentiation with
respect to the redshift parameter. Also, the Hubble rate for the
$\Lambda$CDM model is,
\begin{equation}\label{hubblelcdm}
    H_{\Lambda}(z)=H_0\sqrt{\Omega_{\Lambda} +\Omega_m(z+1)^3 +\Omega_r(z+1)^4 } ,
\end{equation}
where $\Omega_{\Lambda} \simeq 0.68136$ and $\Omega_m \simeq
0.3153$. Moreover, $ H_0 \simeq 1.37187 \times 10^{-33}$eV is the
present day Hubble rate according to the latest 2018 Planck data
\cite{Planck2018}.

\section{Viable $F(R)$ Gravity Models and the de Sitter Scalaron Mass}

In this paper, we analyze viable $F(R)$ theories with different
methods starting from the inflationary era that is assuming to be
a slow-roll era, ending up to a $\Lambda$CDM-like late-time era.
To evaluate the early-time viability of a model we text the
behavior of the parameter \cite{inflation5,Oikonomou:2025qub}
 \begin{equation}\label{xparam}
x= 4\frac{R F_{RRR}}{F_{RR}}\;,
 \end{equation}
 that plays an important role in inflationary and post-inflationary dynamics of a $F(R)$ gravity.
In particular,  all the viable $F(R)$  models which unify early
and late-time acceleration, do yield $-1\leq x\leq 0$
\cite{Oikonomou:2025qub}. It is worth elaborating on this issue
since it is theoretically important. The Einstein frame scalaron
mass of the $F(R)$ is equal,
$$
m^2=\frac{1}{3}\left(-R+\frac{F_R}{F_{RR}} \right)\, ,
$$
which measures the de Sitter perturbations. This can be expressed
in terms of the variable $y$,
$$
m^2=\frac{R}{3}\left(1-\frac{1}{y} \right)\, ,
$$
with $y$ being equal to,
$$
y=\frac{R\,F_{RR}}{F_R}\, .
$$
Requiring that the scalaron mass is either positive or zero, we
obtain the following constraint on the parameter $y$,
$$
0< y \leq 1\, .
$$
Now the important assumption that remarkably produces viable
$F(R)$ gravity models that unify the inflationary era with the
dark energy era, is that the  scalaron mass must be a
monotonically increasing function of $R$ for all the values of the
curvature, thus for both low and high curvatures. This requirement
means that basically, the scalaron mass takes small values in the
small curvature regime-thus at late times- and large values in the
large curvature regime, thus during the inflationary era. The
requirement that the scalaron mass is monotonically increasing in
terms of $R$, that is,
$$
\frac{\partial m^2(R)}{\partial R}\geq 0 \, ,
$$
yields,
$$
\frac{\partial m^2(R)}{\partial
R}=-\frac{1}{3}\frac{F_R}{R\,F_{RR}}\,\frac{R\,F_{RRR}}{F_{RR}}\geq
0 \, ,
$$
or equivalently,
$$
\frac{\partial m^2(R)}{\partial R}=-\frac{1}{3}\frac{x}{y}>0 \, ,
$$
which can be true when,
$$
x\leq 0,\,\,\,0\leq y<1\, .
$$
The models we shall analyze satisfy these requirements, thus can
unify the inflationary era with the dark energy era.

There is an important class of $F(R)$ gravity models which leads
to a unified description of inflation and the dark energy era.
These models have the following simplified form,
\begin{equation}\label{simlifiedexponential}
F(R)=R+\frac{R^2}{M^2}+\lambda
R\,e^{\epsilon\left(\frac{\Lambda}{R}\right)^{\beta}}+\lambda
\Lambda n \epsilon\, ,
\end{equation}
with $\epsilon$, $\lambda$, $\beta$ and $n$ being dimensionless
parameters. This particular class of models yield,
\begin{equation}\label{xexp}
x\sim -\mathcal{C}\,\frac{M^2\Lambda^{\beta}}{R\,R^{\beta}}
\end{equation}
in the large curvature regime during the inflationary era, with
$\mathcal{C}=2\beta  \left(\beta ^2-1\right) \lambda  \epsilon$,
thus $x\sim 0$ and the $R^2$ term dominates the evolution during
the inflationary era. More importantly, these models also yield a
viable dark energy era as we will demonstrate in a later section,
and specifically we will show that $\Omega_{DE}(0)=0.6901$
regarding the dark energy density parameter, while the dark energy
EoS parameter is $\omega_{DE}(0)=-1.036$ for $\beta=0.99$
$\lambda=0.8$, $\epsilon=9.1$ and $n=0.099$. The exceptional class
of exponential deformations of the $R^2$ model stem naturally from
the requirements that the de Sitter mass is a monotonic function
of the Ricci scalar and also that $x$ is almost zero.

Having the above requirements in mind, one can construct viable
$F(R)$ gravity models. Consider for example the model,
\begin{equation}\label{fr22}
    F(R)=R+\frac{R^2}{M^2}-\frac{\beta \Lambda}{c+\log(\epsilon\,R/m_s^2)},
\end{equation}
which primordially is an $R^2$ gravity, and at late-times we
obtain a viable dark energy era by choosing $\beta=0.5\,, c=1 ,
\epsilon=1/220$, and the dark energy era is controlled by the last
term. For this model, we have for example, $\Omega_{DE}(0)=0.6834$
and  $\omega_{DE}(0)=-1.0372$, which are both compatible with the
Planck constraints on the cosmological parameters
$\Omega_{DE}=0.6847 \pm 0.0073$ and $\omega_{DE}=-1.018 \pm
0.031$. This specific model originates from a $x$ parameter which
has the following form,
\begin{equation}\label{xmodel1extra}
x=-\frac{8 \beta  \Lambda  M^2 \left(\log \left(\frac{R \epsilon
}{m_s^2}\right) \left(\log \left(\frac{R \epsilon
}{m_s^2}\right)+5\right)+7\right)}{\left(\log \left(\frac{R
\epsilon }{m_s^2}\right)+1\right) \left(3 \beta  \Lambda M^2+\log
\left(\frac{R \epsilon }{m_s^2}\right) \left(\beta \Lambda M^2+2
R^2 \log \left(\frac{R \epsilon }{m_s^2}\right) \left(\log
\left(\frac{R \epsilon }{m_s^2}\right)+3\right)+6 R^2\right)+2
R^2\right)}\, ,
\end{equation}
and one can easily verify that the parameter $x$ is negative and
$x\sim 0$ during the whole large curvature regime. Another model
of this sort is,
\begin{equation}\label{fr221}
    F(R)=R+\frac{R^2}{M^2}-\frac{\beta  \Lambda }{\gamma +\frac{1}{\log \left(\frac{R \epsilon }{m_s^2}\right)}},
\end{equation}
and as in the previous model, the primordial era is described by
an $R^2$ gravity, and the late-time era the last term dominates,
thus a viable dark energy era is obtained. Specifically, by
choosing $\beta=11.81\,, \gamma=1.5 , \epsilon=100$, we get,
$\Omega_{DE}(0)=0.6876$ and $\omega_{DE}=-0.9891$,  which are both
compatible with the Planck data. In this case, the parameter $x$
takes the form,
\begin{equation}\label{xmodel1extra1}
x=-\frac{8 \beta  \Lambda  M^2 \left(3 \gamma ^2+3 \gamma +\gamma
^2 \log ^2\left(\frac{R \epsilon }{m_s^2}\right)+(3 \gamma +2)
\gamma  \log \left(\frac{R \epsilon
}{m_s^2}\right)+1\right)}{\left(\gamma  \log \left(\frac{R
\epsilon }{m_s^2}\right)+1\right) \left(\beta  (2 \gamma +1)
\Lambda  M^2+\gamma  \left(\beta  \Lambda  M^2+6 R^2\right) \log
\left(\frac{R \epsilon }{m_s^2}\right)+2 \gamma ^3 R^2 \log
^3\left(\frac{R \epsilon }{m_s^2}\right)+6 \gamma ^2 R^2 \log
^2\left(\frac{R \epsilon }{m_s^2}\right)+2 R^2\right)}\, ,
\end{equation}
and in this case, in the large curvature regime, the parameter $x$
is small and negative.

Most of these models contain exponential terms of the Ricci
scalar, which naturally emerge in the formalism due to the
requirement on the values of the parameter $x$. We shall
selectively use some unification $F(R)$ gravity models to confront
them with the latest dark energy observations. Consider for
example the exponential $F(R)$ model
\cite{OdintsovSGS:2017,CognolaENOSZ:2008,Linder:2009}
 \begin{equation} \label{FRexpon}
 F(R)=R -2\Lambda\bigg[1-\exp\bigg(-\varepsilon\frac{R}{2\Lambda}\bigg)\bigg]+
F_\mathrm{inf} \ ,
  \end{equation}
with $\varepsilon$ being a positive constant, and $\Lambda$ being
the cosmological constant. If the Ricci scalar is much larger
compared to $2\Lambda/\varepsilon$ (however $R\ll R_i$ and thus we
can neglect the term responsible for the inflationary era, namely
$F_\mathrm{inf}$) the expression (\ref{FRexpon}) reduces to the
$\Lambda$CDM Lagrangian $F(R)=R -2\Lambda$. The term
$F_\mathrm{inf}={R^2}/{M^2}$ is related with the inflationary
regime,  and the constant $M\sim 3\cdot10^{22}$ eV is assumed to
be large enough, that renders the term $F_\mathrm{inf}$ negligible
during the late-time epoch. Specifically, after the recombination
epoch, at redshifts $0\le z\le10^3$, the Ricci scalar $R$ takes
values less than $5 \cdot 10^{-58}$ eV${}^2$, thus the fraction
$F_\mathrm{inf}/R$ is very small:
  \begin{equation} \label{FinfR}
 \frac{F_\mathrm{inf}}R = \frac{R}{M^2} < 10^{-102}\,.
  \end{equation}
Therefore, we can formally neglect the inflationary term
$F_\mathrm{inf}={R^2}/{M^2}$. In the following we shall use
similar models for our analysis.

The exponential $F(R)$ model (\ref{FRexpon}) was tested on
stability of metric and  matter density perturbations
\cite{Bamba:2012qi,CognolaENOSZ:2008,Linder:2009,ElizaldeNOSZ:2011,CaiS:2014,NunesPSA:2016,Chen:2019},
it satisfies the stability conditions $F_R>0$ and $F_{RR}>0$
during all stages of its evolution. This model is also successful
in late-time observational tests
\cite{OdintsovSGS:2017,OdintsovSGSlog:2019}.

For our formal dark energy analysis,  we rewrite the equation
(\ref{ft10newadd}) or $R=6\dot H + 12H^2$ and in addition the
Friedmann equation (\ref{Friedman}) as follows
\cite{OdintsovSGS:2017,OdintsovSGSlog:2019,OdintsovSGStens:2021}:
 \begin{eqnarray}
\frac{dH}{d\log a}&=&\frac{R}{6H}-2H\ , \label{eqH1} \\
\frac{dR}{d\log
a}&=&\frac1{F_{RR}}\bigg(\frac{\kappa^2\rho}{3H^2}-F_R+\frac{RF_R-F}{6H^2}\bigg)\ .
 \label{eqR1}
 \end{eqnarray}
This system of equations is equivalent to Eq.~(\ref{difyh})
expressed in terms of the statefinder $y_H$ and in addition, the
above system can be integrated numerically for a chosen $F(R)$
model, if we also take into account the following considerations:
(a) one should integrate ``to the future direction'' (with growing
$a$ or decreasing $z$), because in the opposite dynamical variable
direction, that is, ``into the past'', the integral curves of
Eq.~(\ref{difyh}) and the system (\ref{eqH1}), (\ref{eqR1})
diverge, and thus deviate from viable solutions; (b) for the
system (\ref{eqH1}), (\ref{eqR1}) one should carefully define the
initial conditions at some point $a_\mathrm{ini}$ or equivalently
$z_\mathrm{ini}=a_\mathrm{ini}^{-1}-1$ in the past, similarly to
Eqs.(\ref{initialcond}) where $z_\mathrm{ini}\equiv z_f=10$; (c)
for the exponential model (\ref{FRexpon}) and similar $F(R)$
gravity models, the quantity $F_{RR}$ in the denominators of
equations (\ref{difyh}) and (\ref{eqR1}) tends to zero for high
curvature values, that is for large $R$, and therefore, this
uncertainty appearing in the past, needs an accurate approach.
Specifically, for the exponential model (\ref{FRexpon}) the
quantity $F_{RR}$ we mentioned is approximately equal to
 $F_{RR}\simeq\frac{\varepsilon^2}{2\Lambda}\exp\big(-\varepsilon\frac{R}{2\Lambda}\big)$
which is obtained if we neglect the smallest summand $2/M^2$. At
high values of the curvature $R$, $F_{RR}$ in the denominator in
the right hand side of Eq.~(\ref{eqR1}), approaches zero, and thus
we should demand that the numerator vanishes reciprocally too.
This condition is achieved if we use the mentioned fact that the
model (\ref{FRexpon}) approximates $\Lambda$CDM model in the high
curvature $R$ regime or, more precisely, for
$\varepsilon\frac{R}{2\Lambda}\gg1$. Therefore, we assume that in
this regime, and  at earlier times, viable solutions of this
$F(R)$ model asymptotically approach the $\Lambda$CDM model
(\ref{hubblelcdm}) which can be expressed, in the usual way, via
the Hubble constant $H_0$ and the fractions of components,
 \begin{equation} \label{Omega_mL}
\Omega_m=\frac{\kappa^2 \rho_m^{(0)}}{3H_0^2}= \frac{m_s^2}{H_0^2}\,,\qquad
\Omega_\Lambda=\frac{\Lambda}{3H_0^2}\,,\qquad\Omega_r=X_r\Omega_m\;.
 \end{equation}
Nevertheless, at the initial integration point $a_\mathrm{ini}$,
for the Eqs.~(\ref{eqH1}), (\ref{eqR1})  we do not know the Hubble
constant $H_0$, which can be calculated at the end of integration
of the Hubble parameter $H(a)$ at the present epoch $t_0$ (or
$a=1$): $H_0=H(t_0)$. This problem can be solved (a) if we exclude
the usually used cosmological parameters $H_0$, $\Omega_m$, $
\Omega_\Lambda$ and express the asymptotic $\Lambda$CDM solutions
(\ref{hubblelcdm}) using $m_s$ and $\Lambda$ (like we did in the
statefinder parameter $y_H$ approach in the previous section):
 \begin{equation} \label{HLCDM2}
 H_{\Lambda}^2(z)= m_s^2\big[(z+1)^3 +X_r(z+1)^4\big]+ \frac{\Lambda}3\;;
\end{equation}
or (b) we can use preliminary ($\Lambda$CDM approaching) values
$H^{*}_0$ for the Hubble constant, and defined with $H^{*}_0$
parameters,
 \begin{equation} \label{Omega_mL2}
\Omega_m^{*}=\frac{m_s^2}{(H_0^*)^2}\,,\qquad
\Omega_\Lambda^*=\frac{\Lambda}{3(H_0^*)^2}
 \end{equation}
at the initial integration point $a_\mathrm{ini}$ with the
asymptotical $\Lambda$CDM solutions (\ref{hubblelcdm}) or
(\ref{HLCDM2}) for $H(a)$ and the corresponding expression for the
Ricci scalar $R(a)$ recast in the form
\cite{OdintsovSGS:2017,OdintsovSGSlog:2019,OdintsovSGStens:2021,OdintsovSGS_Axi:2023}:
 \begin{equation} \label{asymLCDM}
 \frac{H^2}{H^{*2}_0}=\Omega_m^{*} \big(a^{-3}+ X_r a^{-4}\big)+\Omega_\Lambda^{*}\,,\qquad
 \frac{R}{2\Lambda}=2+\frac{3m_s^2}{2\Lambda}a^{-3}=2+\frac{\Omega_m^{*}}{2\Omega_\Lambda^{*}}a^{-3}\ .
 \end{equation}
In the latter approach, we can obtain the true value of the Hubble
constant $H_0=H(t_0)$, by integrating numerically the system of
equations (\ref{eqH1}), (\ref{eqR1}) from $a_\mathrm{ini}$ to the
present day values $a=1$. Then the parameters $\Omega_m$ and
$\Omega_\Lambda$ can be obtained, from the following relations
originating from Eqs.~(\ref{Omega_mL}) and (\ref{Omega_mL2}), as
follows,
\begin{equation} \label{H0Omm}
 \Omega_m^0H_0^2=\Omega_m^{*}(H^{*}_0)^2=m_s^2\ ,
 \qquad  \Omega_\Lambda H_0^2=\Omega_\Lambda^{*}(H^{*}_0)^2=\frac{\Lambda}3\ .
 \end{equation}
We should note that, the two approaches we described above, are
rather similar: both methods utilize the equivalent equations
(\ref{difyh}) for $y_H(z)$ and (\ref{eqH1}), (\ref{eqR1}) for
$H(a)$, $R(a)$ and similar to the $\Lambda$CDM asymptotical
initial conditions (\ref{asymLCDM}) and (\ref{initialcond}) for
$y_H(z)$. Specifically, for the $\Lambda$CDM model, the
statefinder parameter $y_H(z)$ (\ref{yhdef}) is constant:
 \begin{equation} \label{yHLCDM}
y_H\Big|_{\Lambda\mathrm{CDM}}=\frac\Lambda{3m_s^2}=\frac{\Omega_\Lambda}{\Omega_m}
\end{equation}
 and the initial conditions of Eq. (\ref{initialcond}) at $z_\mathrm{ini}=10$ are very close to this
constant. However, the fixed point $z_\mathrm{ini}=10$ can
restrict the acceptable values of the model parameters, for
example, in the exponential model (\ref{FRexpon}) the values of
$\varepsilon$ are constrained, because the term
$F_{RR}\simeq\frac{\varepsilon^2}{2\Lambda}\exp\big(-\varepsilon\frac{R}{2\Lambda}\big)$
should not be too small at  $z_\mathrm{ini}$. The methodological
approach which utilizes the parameter $y_H(z)$, evaluated in the
range $z\in[0,10]$, encounters another problem: it is necessary to
prolong $H(z)$ to values near the recombination epoch at
$z\sim1100$, if we confront our models with the observational data
including data coming from the CMB and BAO, see a later section
for this discussion. In the next section we thoroughly study three
characteristic examples of new viable $F(R)$ scenarios with the
$\Lambda$CDM asymptotic behavior at the high $R$ limit. Also these
models provide a unified description of the dark energy era and of
the inflationary era. Let us present in brief these models, and
the first of these models, cited below as ``Model I'' is,
\begin{equation}\label{ModFR1}
    F(R)=R+\frac{R^2}{M^2}-\Lambda  \left[\gamma -
    \alpha\exp\left(-\epsilon\frac{R}{m_s^2}\right)\right]\, ,
\end{equation}
is a generalization of the exponential model (\ref{FRexpon})
(reducing to that model if $\gamma=\alpha=2$). We can redefine the
positive constant $\epsilon$ to
$\varepsilon=(2\Lambda/m_s^2)\cdot\epsilon$ and recast
(\ref{ModFR1}) in the equivalent form,
\begin{equation}\label{ModFR11}
    F(R)=R+\frac{R^2}{M^2}-\Lambda  \left[\gamma -
    \alpha\exp\left(-\varepsilon\frac{R}{2\Lambda}\right)\right]\,
    .
\end{equation}
This model primordially behaves as $R^2$ gravity, but at
late-times the inflationary term $F_\mathrm{inf}={R^2}/{M^2}$
becomes negligible. A viable evolution during the dark energy era
may be achieved, by choosing $\gamma=7.5$, $\alpha=1$,
$\epsilon=0.0005$. These values are obtained by solving the
equation (\ref{difyh}) for the statefinder parameter $y_H$ with
the initial conditions (\ref{initialcond}) focusing on the control
for the dark energy density parameter (\ref{OmegaDE}) and  the
dark energy EoS parameter (\ref{EoSDE}). These parameters
$\Omega_{DE}(z)$ and $\omega_{DE}(z)$ at $z=0$ should satisfy the
Planck 2018 constraints \cite{Planck2018}. Regarding the late-time
phenomenology of Model I (\ref{ModFR1}) with the assumed values of
the parameter quoted above, $\gamma$, $\alpha$, $\epsilon$, one
obtains the $\Omega_{DE}(0)=0.6847$  and $\omega_{DE}=-1.0367$,
which are both compatible with the latest Planck data on the
cosmological parameters as we can see in Table~\ref{T1}, where
these estimations are tabulated.

For Model I, the $x$ parameter (\ref{xparam}) takes the quite
simple form:
\begin{equation}\label{xmodel1extra125}
x=-\frac{4\alpha\Lambda  M^2 R \epsilon ^3}{\alpha\Lambda  M^2
m_s^2 \epsilon ^2+2 m_s^6 e^{{\epsilon R}/{m_s^2}}}\, ,
\end{equation}
and it is both very small and negative primordially, as it can be
checked since primordially we have,
 $$
x\sim -\frac{2 \Lambda  M^2 R \epsilon ^3}{m_s^6} e^{-{R \epsilon
}/{m_s^2}}\, ,
 $$ 
thus the model is deemed theoretically viable, since it satisfies
the viability constraints. In the low curvature $R$ limit, the
Model I tends to the $\Lambda$CDM  Lagrangian $F(R)=R-2\Lambda$ if
$\gamma=2$. Values $\gamma\ne2$ for this model are equivalent to
changing the scale of the cosmological constant $\Lambda$. In
further tests we fix $\gamma=2$ for this model. In the nest
section we test this model and in addition two other viable models
which we will present shortly, confronting its predictions with
observational data. For this purpose the system (\ref{eqH1}),
(\ref{eqR1}) is rewritten in the form,
\begin{eqnarray}
\frac{dE}{d\log a}&=&\Omega_\Lambda^{*}\frac{{\cal R}}{E}-2E, \label{eqH2} \\   
\frac{d{\cal R}}{d\log a}&=&2\frac{\big[\Omega_m^{*}(a^{-3}+ X_r a^{-4})
+\Omega_\Lambda^{*}\big(1-\frac12\alpha(1+\varepsilon{\cal R})\,e^{-\varepsilon{\cal
R}}\big)\big]\big/E^2-1+\frac12\alpha\varepsilon e^{-\varepsilon{\cal R}}}
 {\alpha\varepsilon^2 e^{-\varepsilon{\cal R}}}\,,
  \label{eqR2}
  \end{eqnarray}
where we use as variables the normalized Hubble parameter and
Ricci scalar,
\begin{equation}\label{ER}
E=\frac{H}{H_0^{*}},\qquad  {\cal R}=\frac{R}{2\Lambda}\,.
\end{equation}
The initial redshift $z_\mathrm{ini}=a_\mathrm{ini}^{-1}-1$ from
the initial conditions for equations (\ref{eqH2}), (\ref{eqR2}) or
the $\Lambda$CDM asymptotical conditions (\ref{asymLCDM}),  is
determined in the following way: the factor
$\delta=e^{-\varepsilon{\cal R}_\mathrm{ini}}$ in the denominator
of Eq.\~(\ref{eqR2})
 should be should be much smaller than unity. Assuming for the calculations  $\delta\sim
10^{-9}$, one obtains:
\begin{equation}\label{aini}
a_\mathrm{ini}=\bigg[\frac{2\Omega_\Lambda^*}{\Omega_m^*}\bigg(\frac{\log\delta^{-1}}{\varepsilon}-2\bigg)\bigg]^{-1/3}
.
\end{equation}
Let us quote here another viable model (Model II) from the same
class with perplexed form of the parameter $x$, which is the
following,
\begin{equation}\label{fr2212}
    F(R)=R+\frac{R^2}{M^2}-\frac{\beta  \Lambda }{\gamma +\exp \left(- \epsilon\frac{R }{m_s^2}\right)},
\end{equation}
As in the previous model, this model is also primordially
described by an $R^2$ gravity, but at late times, the inflationary
term can be neglected. If the Ricci scalar $R$ is much larger than
$m_s^2/\epsilon$, the function $F(R)$ tends to the $\Lambda$CDM
expression $R-\frac\beta\gamma\Lambda$, and it becomes a pure
$\Lambda$CDM Lagrangian if $\beta=2\gamma$. Using the statefinder
parameter approach we obtain a viable dark energy evolution in
Model II,  by choosing $\beta=20$, $\gamma=2$, $\epsilon=0.00091$.
The corresponding dark energy density parameter and dark energy
EoS parameter are tabulated in Table~\ref{T1}. They are compatible
with the latest Planck data on the cosmological parameters. This
model stems from a $x$ parameter of the form,
\begin{equation}\label{xmodel1extra12}
x=-\frac{4 \beta  \Lambda  M^2 R \epsilon ^3 e^{\frac{R \epsilon }{m_s^2}} \left(\gamma
e^{\frac{R \epsilon }{m_s^2}} \left(\gamma e^{\frac{R \epsilon
}{m_s^2}}-4\right)+1\right)}{m_s^2 \left(\beta \Lambda M^2 \epsilon ^2 e^{\frac{R
\epsilon }{m_s^2}} \left(\gamma ^2 e^{\frac{2 R \epsilon }{m_s^2}}-1\right)+2
\left(\gamma m_s e^{\frac{R \epsilon }{m_s^2}}+m_s\right)^4\right)}\, .
\end{equation}
Now, it can easily be checked that in this case too, the parameter
$x$ is negative and very small, in fact, $x\sim 0$ in the large
curvature regime. Thus the model is theoretically casted in the
viable $F(R)$ gravity models which can also provide a unified
description of inflation and the dark energy era. For Model II, we
can use the same initial condition (\ref{aini}) in the notation
$\varepsilon=\epsilon\Lambda/m_s^2$, because the term $F_{RR}$ in
the denominator is the analog of Eq.\~(\ref{eqR2}) is also
proportional to $e^{-\varepsilon{\cal R}}$. Observational tests
for this model are also described in the next section.

\smallskip

The third
viable $F(R)$ gravity model, which we shall refer to as Model III,
has a peculiar form of the parameter $x$, and it is the following,
\begin{equation}\label{fr22124}
    F(R)=R+\frac{R^2}{M^2}-\frac{\beta  \Lambda  \left(\frac{R}{m_s^2}\right)^n}{\delta +\gamma  \left(\frac{R}{m_s^2}\right)^n},
\end{equation}
As in the previous models, this model is also primordially an
$R^2$ gravity, but at late-times the non-inflationary terms
dominate again, and at large $R$ the Lagrangian tends to the same
expression $R-\frac\beta\gamma\Lambda$ with the power-law rate
instead of exponential law. We shall thorouhghly analyze this
model's late-time phenomenology in the next section. A viable dark
energy era for Model III is accomplished by choosing the
parameters, presented in Table~\ref{T1}. Specifically, regarding
the late-time phenomenology for this model, we get,
$\Omega_{DE}(0)=0.6851$ regarding the dark energy density
parameter, while the dark energy EoS parameter is
$\omega_{DE}=-0.9887$,  which are again compatible with the Planck
data on the cosmological parameters. This model stems from a $x$
parameter of the form,
\begin{equation}\label{xmodel1extra124}
x=\frac{4 \beta  \delta  \Lambda  M^2 n \left(\frac{R}{m_s^2}\right)^n \left(4 \gamma
\delta \left(n^2-1\right) \left(\frac{R}{m_s^2}\right)^n-\gamma ^2 (n+1) (n+2)
\left(\frac{R}{m_s^2}\right)^{2 n}-\delta ^2 (n-2) (n-1)\right)}{\beta  \delta  \Lambda
M^2 n \left(\frac{R}{m_s^2}\right)^n \left(\delta +\gamma
\left(\frac{R}{m_s^2}\right)^n\right) \left(\delta +\gamma (n+1)
\left(\frac{R}{m_s^2}\right)^n-\delta  n\right)+2 R^2 \left(\delta +\gamma
\left(\frac{R}{m_s^2}\right)^n\right)^4}\, .
\end{equation}
Now it can also be checked that in this case too, the parameter
$x$ negative and very small, and in fact, $x\sim 0$ in the large
curvature regime. Hence, all the three viable models we quoted
above, namely Model I-III, result to a unification of early and
late-time acceleration, and also all these models yield
primordially $x$ in the range $-1\leq x\leq 0$, and in fact $x\sim
0$ and negative \cite{Oikonomou:2025qub}. Hence these are
theoretically viable models. In the next section we shall
thoroughly investigate the late-time viability of Models I, II and
III confronting these with the latest observational data.
\begin{table}[h!]
  \begin{center}
    \caption{
    Cosmological Parameters Values at present day for the models (\ref{ModFR1}), (\ref{fr2212})  and (\ref{fr22124}).}
    \label{T1}
    \begin{tabular}{|c|c|c|c|c|}
    \hline
      \textbf{Parameter} & Model I (\ref{ModFR1})& Model II (\ref{fr2212}) &  Model III (\ref{fr22124}) & \textbf{Planck 2018}
      \\  \hline
      $\Omega_{DE}(0)$ & $0.6847$  & $0.6918$ & 0.6851 & $0.6847\pm 0.0073$ \\  \hline
      $\omega_{DE}(0)$ & $-1.0367$ & $-0.9974$ & -0.98876 & $-1.018\pm0.031$       \\  \hline
      model & $\gamma=7.5$, $\alpha=1$,&$\beta=20,\;\gamma=2$ &$\beta=1.4\,,\gamma=0.2$, & -       \\
      parameters &  $\epsilon=0.0005$   &$\epsilon=0.00091$ & $\delta=0.2,\,n=0.3$ & -
      \\  \hline
    \end{tabular}
  \end{center}
\end{table}
Now a good question is whether the $F(R)$ gravity terms that
determine the late-time era in the models I-III above actually
affect the inflationary era. This is a good question, which was
addressed in Ref. \cite{Oikonomou:2025qub}. Specifically, using a
model agnostic approach, it was shown in \cite{Oikonomou:2025qub}
that the spectral index of the primordial scalar perturbations for
any $F(R)$ gravity reduces to,
\begin{equation}\label{asxeto1}
n_s-1=-4\epsilon_1+x\epsilon_1\, ,
\end{equation}
and the tensor-to-scalar ratio is equal to,
\begin{equation}\label{mainequation}
r\simeq \frac{48 (1-n_s)^2}{(4-x)^2}\, .
\end{equation}
Hence, for all the models I-III we examined earlier, the
inflationary era is dominated by an $R^2$ term because
primordially we have approximately $x\sim 0$ for all these models.
But this can be seen easily by the leading order behavior of the
$F(R)$ gravity functions in Eqs. (\ref{ModFR1}), (\ref{fr2212})
and (\ref{fr22124}). As it can be seen, in the large curvature
regime, only the $R^2$ term dominates.

\section{Confronting the $F(R)$ Gravity Models with the Observational Data: A Thorough Statistical Analysis}
\label{Observ}

The suggested   models of the previous section, namely
(\ref{ModFR1}), (\ref{fr2212})  and (\ref{fr22124}) (Models I, II
and III) should be confronted with up-to-date observational data.
We also compare them in these tests with the exponential $F(R)$
model (\ref{FRexpon}) and with the $\Lambda$CDM scenario
(\ref{hubblelcdm}). Any new viable model should demonstrate some
advantages. We include in our tests the following observational
data: (a)  Type Ia Supernovae (SNe Ia) data from the Pantheon+
sample database \cite{PantheonP:2022}, (b) estimations of the
Hubble parameter $H(z)$ or Cosmic Chronometers (CC), (c)
parameters from the Cosmic Microwave Background radiation (CMB)
and the recent Baryon Acoustic Oscillations (BAO) data from Dark
Energy Spectroscopic Instrument (DESI) collaboration
\cite{DESI:2024,DESI:2025zgx}. For SNe Ia data the Pantheon+
catalogue \cite{PantheonP:2022} is used, which provides
$N_{\mathrm{SN}}=1701$ datapoints that contains information of the
distance moduli $\mu_i^\mathrm{obs}$ at redshifts $z_i$ from 1550
SNe Ia. For a tested cosmological model with a set of its free
parameters $\theta_1,\theta_2,\dots$ we determine the Hubble rate
$H(z)$ and like in the previous paper \cite{OdintsovSGS:2024}
calculate the $\chi^2$ function:
\begin{equation}
\chi^2_{\mathrm{SN}}(\theta_1,\dots)=\min\limits_{H_0} \sum_{i,j=1}^{N_\mathrm{SN}}
 \Delta\mu_i\big(C_{\mathrm{SN}}^{-1}\big)_{ij} \Delta\mu_j\ ,\qquad 
 \Delta\mu_i=\mu^\mathrm{th}(z_i,\theta_1,\dots)-\mu^\mathrm{obs}_i\ .
 \label{chiSN}\end{equation} 
Here $C_{\mbox{\scriptsize SN}}$ is the $N_{\mathrm{SN}}\times
N_{\mathrm{SN}}$ covariance matrix and $\mu^\mathrm{th}$ are the
theoretical estimates for the distance moduli:
\begin{equation}
 \mu^\mathrm{th}(z) = 5 \log_{10} \frac{(1+z)\,D_M(z)}{10\mbox{pc}},\qquad D_M(z)= c \int\limits_0^z\frac{d\tilde z}{H(\tilde
 z)}.    \label{muDM}
\end{equation}
As the Hubble parameter data $H(z)$ we include here in our
analysis $N_H=32$ datapoints of $H^\mathrm{obs}(z_i)$ with
references in the previous papers
\cite{OdintsovSGS_Axi:2023,OdintsovSGS_LnAx:2024,OdintsovSGS:2024}.
These datapoints named ``Cosmic Chronometers'' (CC) and are
measured by the method of differential ages $\Delta t$ for
galaxies with known variations of redshifts $\Delta z$ via the
relation: $H (z)= {\dot{a}}/{a} \simeq -\frac{1}{1+z} \frac{\Delta
z}{\Delta t}$.  The $\chi^2$ function for these $H(z)$ estimations
yields:
\begin{equation}
\chi^2_{H}= \sum_{i=1}^{N_H} \left[\frac{H^\mathrm{obs}(z_i)
 -H^\mathrm{th}(z_i; \theta_k)}{\sigma_{H,i}}\right]^2 \, .
\label{chiH}
\end{equation}

All the mentioned above SNe Ia and  $H(z)$ datapoints are measured
in the redshift range $0<z<2.4$, but the CMB observational
parameters are related to the photon-decoupling epoch at redshifts
near $z_*=1089.80 \pm0.21$ and are used here as the set
\cite{Planck2018},
\begin{equation}
\mathbf{x}=\left(R,\ell_A,\omega_b \right)\, ,\quad
R=\sqrt{\Omega_m^0}\frac{H_0D_M(z_*)}c\, ,\quad \ell_A=\frac{\pi D_M(z_*)}{r_s(z_*)}\, ,
\quad\omega_b=\Omega_b^0h^2
 \label{CMB} \end{equation}
in Planck 2018 data with estimations from Ref.~\cite{ChenHW:2018}:
\begin{equation}
\mathbf{x}^\mathrm{Pl}=\left( R^\mathrm{Pl},\ell_A^\mathrm{Pl},\omega_b^\mathrm{Pl}
\right) =\left( 1.7428\pm0.0053,\;301.406\pm0.090,\;0.02259\pm0.00017 \right) \, .
\label{CMBpriors}
\end{equation}
These data priors are adopted for scenarios with zero spatial
curvature and with $\Lambda$CDM-like asymptotic behavior.

The comoving sound horizon  $r_s(z_*)$ is calculated as the
integral
\cite{OdintsovSGS_Axi:2023,OdintsovSGS_LnAx:2024,OdintsovSGS:2024}:
  \begin{equation}
r_s(z)=  \int_z^{\infty} \frac{c_s(\tilde z)}{H (\tilde z)}\,d\tilde
z=\frac1{\sqrt{3}}\int_0^{1/(1+z)}\frac{da}
 {a^2H(a)\sqrt{1+\big[3\Omega_b^0/(4\Omega_\gamma^0)\big]a}}\ ,
  \label{rs2}\end{equation}
where $z_*$ is estimated following
Refs.~\cite{OdintsovSGS_Axi:2023,ChenHW:2018}. The reduced baryon
fraction $\omega_b$ is considered as the nuisance parameter in the
following $\chi^2$ function:
\begin{equation}
\chi^2_{\mathrm{CMB}}=\min_{\omega_b,H_0}\Delta\mathbf{x}\cdot
C_{\mathrm{CMB}}^{-1}\left( \Delta\mathbf{x} \right)^{T}\, ,\quad \Delta
\mathbf{x}=\mathbf{x}-\mathbf{x}^\mathrm{Pl}\,, \label{chiCMB}
\end{equation}
where $C_{\mathrm{CMB}}=\| \tilde C_{ij}\sigma_i\sigma_j \|$ is  the covariance matrix
\cite{ChenHW:2018}. \\
For the Baryon Acoustic Oscillations (BAO) we consider new data
from Dark Energy Spectroscopic Instrument (DESI) from Data Release
1 \cite{DESI:2024} (DR1, 2024) and the latest  Data Release 2
\cite{DESI:2025zgx} (DR2, 2025).  We calculate and compare with
measurements the value,
\begin{equation}
d_z^{-1}(z)= \frac{D_V(z)}{r_s(z_d)}\, ,\qquad D_V(z)=\bigg[\frac{cz D_M^2(z)}{H(z)}
\bigg]^{1/3}\, , \label{dzDV}
\end{equation}
with $z_d$ being the redshift at the end of the baryon drag era,
whereas the comoving sound horizon $r_s(z)$ is calculated as the
integral (\ref{rs2}). The estimations for $z_d$ and for the baryon
to photon ratio $\Omega_b^0/\Omega_\gamma$ are fixed by the Planck
2018 data \cite{Planck2018}.

In this paper, we use  BAO data, shown in Table~\ref{DESI} and
provided by DESI DR1 \cite{DESI:2024} and  DESI DR2
\cite{DESI:2025zgx} with 7 and 8 datapoints respectively. These
measurements include BAO data from clustering of galaxies,
including  ''bright galaxy sample'' (BGS), luminous red galaxies
(LRG), emission line galaxies (ELG), quasars and the
Lyman-$\alpha$ forest in the redshift range $0.1<z<4.2$. The
$\chi^2$ function is,
\begin{equation}
\chi^2_{\mathrm{BAO}}(\theta_1,\dots)=\sum_{i=1}^{N_\mathrm{BAO}}
\left[\frac{d_z^\mathrm{obs}(z_i)-d_z^\mathrm{th}(z_i,\dots)}{\sigma_{d_z,i}}\right]^2
\, . \label{chiBAO}
\end{equation}
\begin{table}[ht]
\begin{center}\label{DESI}
\caption{DESI DR1 and DR2 BAO data for $D_V(z)/r_s(z_d)$.}
\begin{tabular}{||l|c|c|c||c|c||}\hline
 Tracer &$z_\mathrm{eff}$ & $z$ range &  DR1 $D_V/r_d$ &$z_\mathrm{eff}$ & DR2 $D_V/r_d$ \\
\hline
 BGS & 0.295 &0.1 - 0.4 &$7.93\pm 0.15 $& 0.295&$7.942\pm0.075 $ \\
\hline
 LRG1 & 0.51 &0.4 - 0.6 &$12.563\pm0.282 $& 0.51 &$12.720 \pm0.099$ \\
\hline
 LRG2 & 0.706&0.5 - 0.8 &$ 15.898\pm0.354 $ & 0.706 &$16.050 \pm0.110$\\
\hline%
 LRG3 & 0.93&0.8 - 1.1  & $19.865\pm0.315$& 0.922 & $19.656\pm0.105$ \\
 ELG1 &     &           &                 & 0.955 & $20.008\pm0.183$ \\ \hline
 ELG2 &1.317& 1.1 - 1.6 &$24.13\pm0.63$& 1.321 & $24.252\pm0.174$ \\ \hline
 QSO & 1.491&0.8 - 2.1  & $26.07\pm0.67$& 1.484 & $26.055\pm0.398$ \\ \hline
 Ly$\alpha$ & 2.33&1.77 - 4.2  & $31.516\pm0.73$& 2.33 & $31.267\pm0.256$ \\ \hline
 \end{tabular}
\end{center}
\end{table}
Then, the free parameters for the considered models should be
fitted with all these distinct observational data. To estimate
viability of any scenario in this test we calculate the total
$\chi^2$ function with the contributions from SNe Ia, CC, CMB and
BAO DESI with DR1 and DR2 is computed:
 \begin{equation}
  \chi^2=\chi^2_\mathrm{SN}+\chi^2_H+\chi^2_\mathrm{CMB}+\chi^2_\mathrm{BAO}\ .
 \label{chitot} \end{equation}

In our numerical calculations we use the methods developed in the
previous papers
\cite{OdintsovSGS:2024,Odintsov:2025kyw,OdintsovSGS_Axi:2023,OdintsovSGS_LnAx:2024}.
For each set of $N_p$ model free parameters we calculate this
$\chi^2$  function, for each pair of chosen parameters we search
the minimum of parameters. Grid spacing and size of the box are
determined at the initial stage of calculations, the center of the
box is corrected during this process with approximation of the
previous minima. In this approach the prior ranges for the model
parameters are rather wide and determined from their physical
sense, in particular, for Models I and II they are
 $$
 \Omega_m\in[0.1,0.5];\;\;  \Omega_\Lambda\in[0.4,1];\;\; \varepsilon\in[0.01,6];\;\; \alpha\in[0.1,5];\;\;  \gamma\in[0.1,10];\;\;
 H_0\in[50,100]\;\mathrm{km/s/Mpc}\,.
$$

  \begin{figure}[th]
   \centerline{ \includegraphics[scale=0.68,trim=5mm 0mm 2mm -1mm]{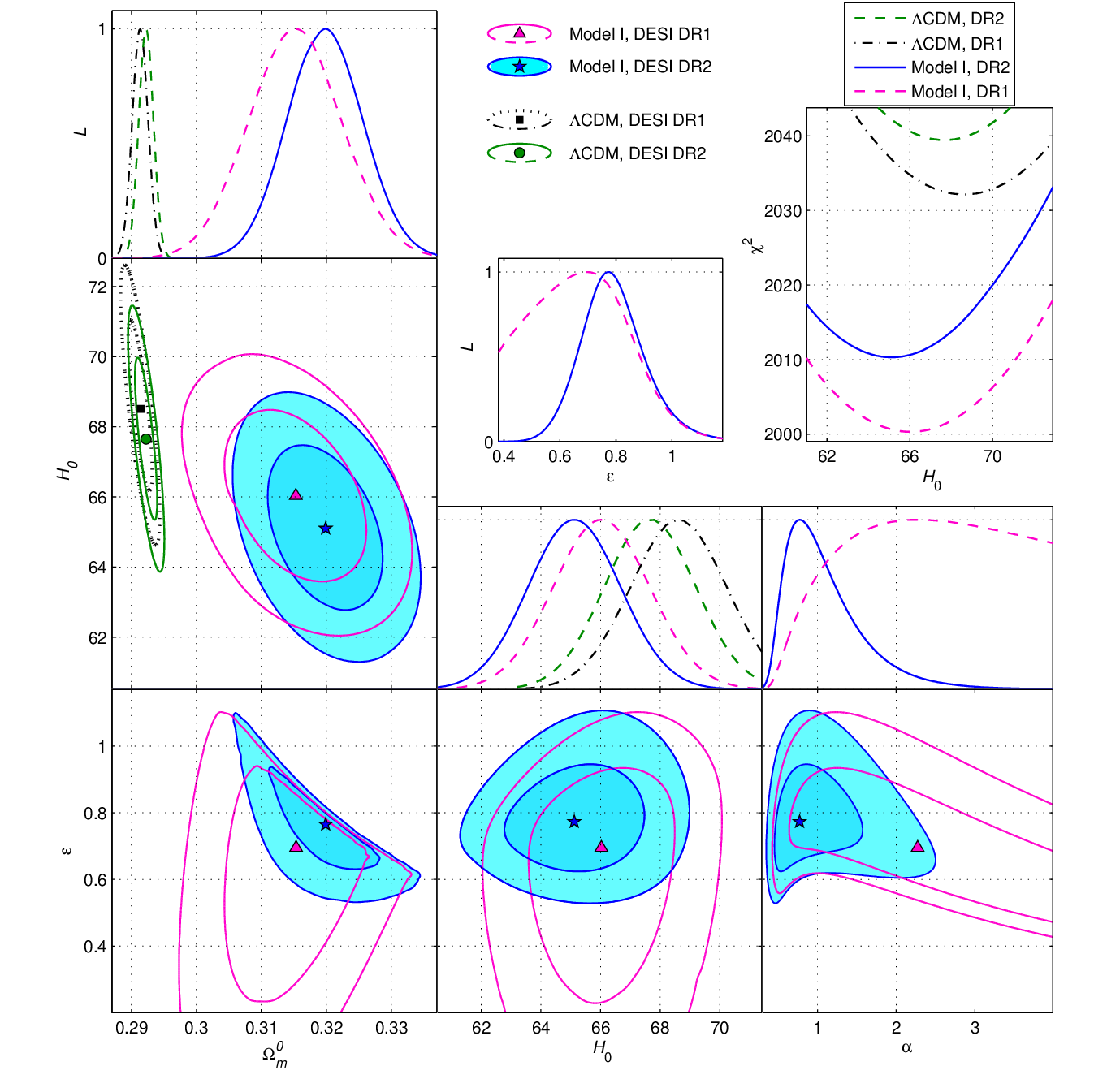}}
\caption{Contour plots of $\chi^2$ with $1\sigma$, $2\sigma$ CL, likelihood functions $
{\cal L}(\theta_i)$ and one-parameter distributions $\chi^2(H_0)$ for Model I
(\ref{ModFR11}) in comparison with and $\Lambda$CDM models for SNe Ia, CC, CMB and two
variants of BAO DESI data: DR1 and DR2. }
  \label{F2}
\end{figure}

We begin our analysis with Model I (\ref{ModFR11}) and compare its minimum of the total
$\chi^2$ function (\ref{chitot}) and the best fits for its model parameters $\alpha$,
$\varepsilon$, $\Omega_m$, $\Omega_\Lambda$, $H_0$ with the similar predictions of the
exponential $F(R)$ model (\ref{FRexpon}) and the $\Lambda$CDM scenario
(\ref{hubblelcdm}). We also compare these results for two variants of BAO DESI data: DR1
and DR2 (see Table~\ref{DESI}), illustrate them in Fig.~\ref{F2} and tabulate
$\min\chi^2$ and the best fits of model parameters in Table~\ref{TMod1}. The contour
plots in $\Omega_m-H_0$, $\Omega_m-\varepsilon$, $H_0-\varepsilon$ $\alpha-\varepsilon$
planes in Fig.~\ref{F2} correspond to $1\sigma$ (68.27\%) and $2\sigma$ (95.45\%)
confidence regions for the two-parameter distributions $\chi^2(\theta_i,\theta_j)$,
which are results of minimizing the $\chi^2$ over all the remaining free parameters. For
example, the contours depicted in the $\Omega_m-H_0$ panel of Fig.~\ref{F2} for Model I
are obtained by
 $$\chi^2(\Omega_m,H_0)=\min\limits_{\alpha,\beta,\Omega_\Lambda}\chi^2(\alpha,\dots,H_0)\;.$$

The stars, circles and other symbols denote the best fits with $\min\chi^2$ of the
corresponding two-dimensional distributions. The best fits with the corresponding
$1\sigma$ errors for the free model parameters  are also shown in Table~\ref{TMod1}
below. These values also can be seen in Fig.~\ref{F2} in the one-parameter distributions
$\chi^2(H_0)$  and the likelihoods ${\cal L}(\theta_j)$ for model parameters $\theta_j$,
which are calculated as:
   \begin{equation}
{\cal L}(\theta_j)= \exp\bigg[- \frac{\chi^2(\theta_j)-m^\mathrm{abs}}2\bigg]\ ,
 \label{likeli} \end{equation}
where $\chi^2(\theta_j)=\min\limits_{\mbox{\scriptsize other
}\theta_k}\chi^2(\theta_1,\dots)$ and $m^\mathrm{abs}$ the absolute minimum for
$\chi^2$. As shown in Table~\ref{TMod1} and in the top-right panel with  $\chi^2(H_0)$
plots in Fig.~\ref{F2}, Model I  (\ref{ModFR11}) achieves the best results in
$m^\mathrm{abs}=\min\chi^2$ both for two considered variants of observational data: for
BAO DESI DR1 $m^\mathrm{abs}\simeq 2000.31$ and for DR2 $m^\mathrm{abs}\simeq 2010.30$.
These results for both DESI DR1 and DR2 are much better than the correspondent
$\min\chi^2$ values of the $\Lambda$CDM model. This advantage does not vanish even when
considering the number of free parameters $N_p$ for each model following the Akaike
information criterion \cite{Liddle_ABIC:2007}
 \begin{equation}
 \mbox{AIC} = \min\chi^2 +2N_p.
  \label{AIC}\end{equation}

 \begin{figure}[th]
   \centerline{ \includegraphics[scale=0.69,trim=5mm 0mm 2mm -1mm]{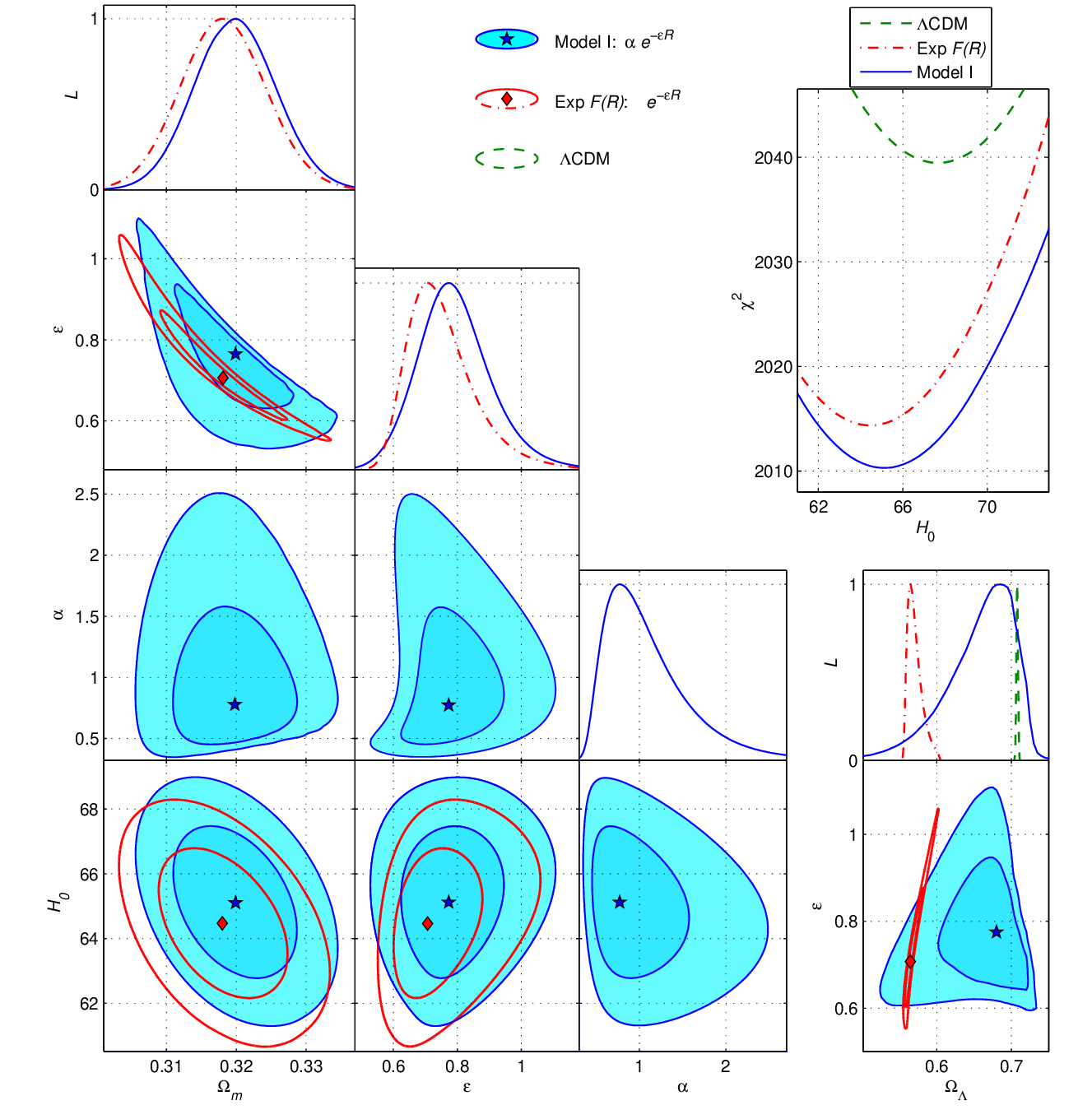}}
\caption{Model I (\ref{ModFR11}) compared with the exponential $F(R)$ model
(\ref{FRexpon}) in  $1\sigma$ and $2\sigma$ contour plots of $\chi^2$, likelihoods $
{\cal L}(\theta_i)$ and one-parameter distributions $\chi^2(H_0)$ for SNe Ia, CC, CMB
and BAO DESI DR2 data. }
  \label{F3}
\end{figure}

Model I has $N_p=5$ free parameters $\alpha$, $\varepsilon$, $\Omega_m$,
$\Omega_\Lambda$, $H_0$ related with our tests, it is noticeably  larger than two
parameters $\Omega_m$ and $H_0$ of $\Lambda$CDM model. The number $N_p=5$ for Model I
(and $N_p=4$ for the exponential model) adds the penalty in AIC, but the resulting AIC
for the alternative models appears to be essentially lower than that for $\Lambda$CDM.
To emphasize this fact the difference
$\Delta\mbox{AIC}=\mbox{AIC}_\mathrm{model}-\mbox{AIC}_{\Lambda\mathrm{CDM}}$ is
included in Table \ref{TMod1}.
\begin{table}[ht]
\begin{center}
\caption{Best fits with $1\sigma$ errors, $\min\chi^2$, AIC and
$\Delta$AIC from SNe Ia, $H(z)$,  CMB and BAO DESI DR1 in
comparison with DR2 datasets for Model I (\ref{ModFR11}),
$\Lambda$CDM and the exponential model (\ref{FRexpon}).}
 \begin{tabular}{|l|c|c|c|c|c|c|c|c|}  \hline
 \hline  Model & DESI &  $\min\chi^2/d.o.f$& AIC & $\Delta$AIC& $\Omega_m$& $H_0$& $\varepsilon$& $\alpha$ \\
\hline\hline
 Model I & DR1 & 2000.31 /1738 & 2010.31&$-25.84$ &$0.3153^{+0.0070}_{-0.0072}$
& $66.03^{+1.62}_{-1.61}$ & $0.695^{+0.166}_{-0.281}$ &$2.27^{+4.50}_{-1.36}$  \rule{0pt}{1.1em}  \\
\hline
$\Lambda$CDM& DR1 & 2032.15 /1741 & 2036.15& 0 & $0.2913^{+0.0013}_{-0.0012}$& $68.60^{+1.62}_{-1.58}$& - & -   \rule{0pt}{1.1em}  \\
\hline
 Exp\,$F(R)$ & DR1 & 2000.32 /1739 & 2008.32&$-27.83$& $0.3158^{+0.0061}_{-0.0058}$ & $66.05^{+1.58}_{-1.63}$ & $0.721^{+0.101}_{-0.078}$ &- \rule{0pt}{1.1em}  \\
\hline \hline
 Model I & DR2 & 2010.30 /1739 & 2020.30& $-23.16$ &$0.3199^{+0.0057}_{-0.0060}$
& $65.12^{+1.55}_{-1.55}$ & $0.773^{+0.106}_{-0.096}$ &$0.774^{+0.455}_{-0.243}$  \rule{0pt}{1.1em}  \\
\hline
$\Lambda$CDM& DR2 & 2039.46 /1742 & 2043.46& 0 & $0.2923^{+0.0011}_{-0.0012}$& $67.65^{+1.55}_{-1.62}$& - & -   \rule{0pt}{1.1em}  \\
\hline
 Exp\,$F(R)$ & DR2 & 2014.36 /1740 & 2022.36& $-21.10$ &$0.3180^{+0.0061}_{-0.0060}$
& $64.46^{+1.53}_{-1.52}$ & $0.707^{+0.102}_{-0.075}$ &- \rule{0pt}{1.1em}  \\
\hline
  \hline \end{tabular}
\end{center}\label{TMod1}
\end{table}

Large negative values $\Delta\mbox{AIC}$ for Model I (\ref{ModFR11}) show its more than
3$\sigma$ advantageous in comparison to  $\Lambda$CDM model, this  advantage is kept for
both variants DR1 and DR2 of BAO DESI data. The exponential $F(R)$ model (\ref{FRexpon})
also demonstrates large negative $\Delta\mbox{AIC}$ in Table~\ref{TMod1}, its behavior
is illustrated in Fig.~\ref{F3} below. In Ref.~\cite{OdintsovSGS:2024} we concluded that
the advantage of $F(R)$ models in comparison to $\Lambda$CDM model in $\min\chi^2$ and
AIC is connected mainly with the last Pantheon+ SNe Ia data \cite{PantheonP:2022}. For
the previous SNe Ia catalogue Pantheon sample 2017 $F(R)$ scenarios always conceded to
$\Lambda$CDM model in AIC
\cite{OdintsovSGS:2017,OdintsovSGSlog:2019,OdintsovSGStens:2021,OdintsovSGS_Axi:2023}.

Large negative  AIC differences in Table~\ref{TMod1} can be achieved not only for $F(R)$
scenarios, we obtained similar results also for $w$CDM model and $w_0w_a$CDM model with
dynamical DE \cite{OdintsovSGS:2024} and for Einstein-Gauss-Bonnet gravity in
Ref.~\cite{Odintsov:2025kyw}. The advantage of different models over $\Lambda$CDM coming
from new SNe Ia and DESI DR2 BAO data was seen in many recent papers
\cite{vanderWesthuizen:2025iam,Paliathanasis:2025dcr,Pan:2025qwy,Yang:2025mws}. In
particular, in Ref.~\cite{Pan:2025qwy} the difference $\Delta\mbox{AIC}=-56.81$ was
obtained for the scenario with interacting dark components and observations including
Pantheon+ and DESI DR2 BAO data. However, some models are not so successful, for
example, the Barrow\,--\,Tsallis holographic dark energy scenario in
Ref.~\cite{Luciano:2025elo} does not exhibit an advantage in AIC over the $\Lambda$CDM
model.

From Table~\ref{TMod1} and Fig.~\ref{F2} one may conclude that two variants of DESI BAO
data do not change the sign of $\Delta\mbox{AIC}$, but essentially influence on AIC and
the best fits of the model parameters. For example, when we use in DR2 the more
stringent limits for $D_V/r_d$ data in comparison with DR1 (see Table~\ref{DESI}), the
best fits for $\alpha$ for Model I appeared to be shifted from
$\alpha=2.27^{+4.50}_{-1.36}$ to $\alpha=0.774^{+0.455}_{-0.243}$. At $\alpha=2$ Model I
transforms into the exponential $F(R)$ model (\ref{FRexpon}), so for DR1 these two
models show the close values of $\min\chi^2$, but the exponential model wins in AIC
because of smaller $N_p$. However for DESI DR2 data lower values of $\alpha$ become
preferable and Model I is advantageous over the exponential model not only in
$\min\chi^2$ but also in AIC. For Model I we also can see in Fig.~\ref{F2} different
behavior of $1\sigma$ and $2\sigma$ CL domains for DR1 and DR2 variants  of DESI BAO
data: DR1 permits sets of model parameters with $\varepsilon<0.5$ and $\alpha>2.5$, but
DR2 excludes these values from the mentioned domains. In the $\Omega_m-H_0$ plane in
Fig.~\ref{F2} one can see that the different behavior of Model I and $\Lambda$CDM
scenario leads not only to the large $\Delta$AIC, but also to different predictions for
the best fits of the Hubble constant $H_0$ and for the matter density parameter
$\Omega_m$ for both variants of BAO data. From Table~\ref{TMod1} for DR2 BAO data  the
$\Lambda$CDM best fit of the Hubble constant is given by $H_0=67.65^{+1.55}_{-1.62}$
km/(s$\cdot$Mpc) whereas for Model I (\ref{ModFR11}), it leads to
$H_0=65.12^{+1.55}_{-1.55}$ km/(s$\cdot$Mpc) with about  $1\sigma$ difference. For the
matter density parameter $\Omega_m$ the $\Lambda$CDM  and $F(R)$ Model I are excluded to
more than $3\sigma$ in their predictions. In our further analysis we concentrate on the
latest and more exact variant DR2 of DESI BAO data. For this variant we compare in
detail  Model I (\ref{ModFR11}) and  the exponential $F(R)$ model (\ref{FRexpon}) in
Fig.~\ref{F3} and Table~\ref{Estim}.

In  Fig.~\ref{F3} one can see the difference between Model I and
the exponential $F(R)$ model in their best fits of common
parameters: Model I predicts slightly enlarged estimates for
$\varepsilon$, $\Omega_m$, $H_0$ and essentially larger value for
$\Omega_\Lambda$ with enhanced 1$\sigma$ error box. The top-right
panel of  Fig.~\ref{F3} with plots $\chi^2(H_0)$ illustrates the
bad result of $\Lambda$CDM scenario and the mentioned advantage of
Model I in competition with the exponential model in $\min\chi^2$
for  DR2 DESI BAO data. This success of Model I is connected with
the best fit $\alpha=0.774^{+0.455}_{-0.243}$ that is far from
$\alpha=2$ that reduces Model I to the exponential $F(R)$ model.
Model I is also more successful in Akaike information criterion
(\ref{AIC}). However this advantage vanishes because of large
number of data point $N_d=1744$, if we  consider the Bayesian
information criterion (BIC) \cite{Liddle_ABIC:2007}
  \begin{equation}
\mathrm{BIC} = \min\chi^2 +N_p\cdot\log(N_d)\;.
 \label{BIC} \end{equation}
In Table~\ref{Estim} one can see that BIC diminishes the distance
between $\Lambda$CDM model and other more successful scenarios,
but  $\Lambda$CDM remains the last model also in BIC competition.
In Table~\ref{Estim} we included the results of similar
calculations for Model II (\ref{fr2212}). For convenience we use
the common for other models parameter
$\varepsilon=2\epsilon\Lambda/m_s^2$ and rewrite the Lagrangian
(\ref{fr2212}) for case $\beta=2\gamma$ (corresponding to the
$\Lambda$CDM limit at high $R$) in the equivalent form
\begin{equation}\label{ModFR2}
    F(R)=R+\frac{R^2}{M^2}-\frac{2\gamma\Lambda }{\gamma +\exp \left(- \varepsilon\frac{R }{2\Lambda}\right)}.
\end{equation}
In this notation Model II has the same number $N_p=5$ of free parameters as Model I with
$\gamma$ instead of $\alpha$. It is not reduced to the  exponential $F(R)$ model, but
can be described by the system including Eq.~(\ref{eqH2}) and the following analog of
equation (\ref{eqR2}):
$$
\frac{d{\cal R}}{d\log a}=\gamma_R\frac{\big[\Omega_m^{*}(a^{-3}+ X_r a^{-4})\gamma_R^2
+\Omega_\Lambda^{*}\gamma\big(\gamma+(1-\varepsilon{\cal R})\,e^{-\varepsilon{\cal
R}}\big)\big]\big/E^2-\gamma_R^2+\gamma\varepsilon e^{-\varepsilon{\cal R}}}
 {\gamma\varepsilon^2 e^{-\varepsilon{\cal R}}\,(\gamma-e^{-\varepsilon{\cal R}})}\,,
$$
 where $\gamma_R=\gamma+e^{-\varepsilon{\cal R}}$. As a starting point for calculations
we use the $\Lambda$CDM asymptotical initial conditions
(\ref{asymLCDM}) at the same initial point $a_\mathrm{ini}$
(\ref{aini}) following from the small value of the factor
$\delta=e^{-\varepsilon{\cal R}_\mathrm{ini}}$ in $F_{RR}$.

\begin{table}[ht]
\begin{center}
\caption{Best fits, $\min\chi^2$, AIC, BIC from SNe Ia, $H(z)$, CMB and BAO DESI DR2
datasets for for Model I (\ref{ModFR11}), Model II (\ref{fr2212}), the exponential
(\ref{FRexpon}) and $\Lambda$CDM models.}
\begin{tabular}{|l|c|c|c|c|c|c|c|c|c|}  \hline
 \hline  Model &   $\min\chi^2/d.o.f$& AIC & BIC& $\Omega_m$& $H_0$& $\Omega_\Lambda$&  $\varepsilon$& $\alpha$ & $\gamma$ \\
\hline
 Model I & 2010.30 /1739 & 2020.30& $2047.62$ &$0.3199^{+0.0057}_{-0.0060}$
& $65.12^{+1.55}_{-1.55}$  & $0.685^{+0.028}_{-0.048}$ &$0.773^{+0.106}_{-0.096}$ &$0.774^{+0.455}_{-0.243}$ & -\rule{0pt}{1.1em}  \\
\hline
 Model II & 2010.23 /1739 & 2020.23&$2047.55$&$0.3194^{+0.0058}_{-0.0052}$
& $65.07^{+1.55}_{-1.56}$ & $0.690^{+0.027}_{-0.044}$ &$0.857^{+0.103}_{-0.103}$ & - &$2.210^{+1.109}_{-0.886}$ \rule{0pt}{1.1em}  \\
\hline
 Exp\,$F(R)$ & 2014.36 /1740 & 2022.36& $2044.22$ &$0.3180^{+0.0061}_{-0.0060}$
& $64.46^{+1.53}_{-1.52}$  & $0.5645^{+0.010}_{-0.006}$ &$0.707^{+0.102}_{-0.075}$ &- & - \rule{0pt}{1.1em}  \\
\hline
$\Lambda$CDM& 2039.46 /1742 & 2043.46& 2054.39 & $0.2923^{+0.0011}_{-0.0012}$& $67.65^{+1.55}_{-1.62}$& - & - & - & - \rule{0pt}{1.1em}  \\
\hline
  \hline \end{tabular}
 \end{center} \label{Estim}
\end{table}

\begin{figure}[th]
   \centerline{ \includegraphics[scale=0.69,trim=5mm 0mm 2mm -1mm]{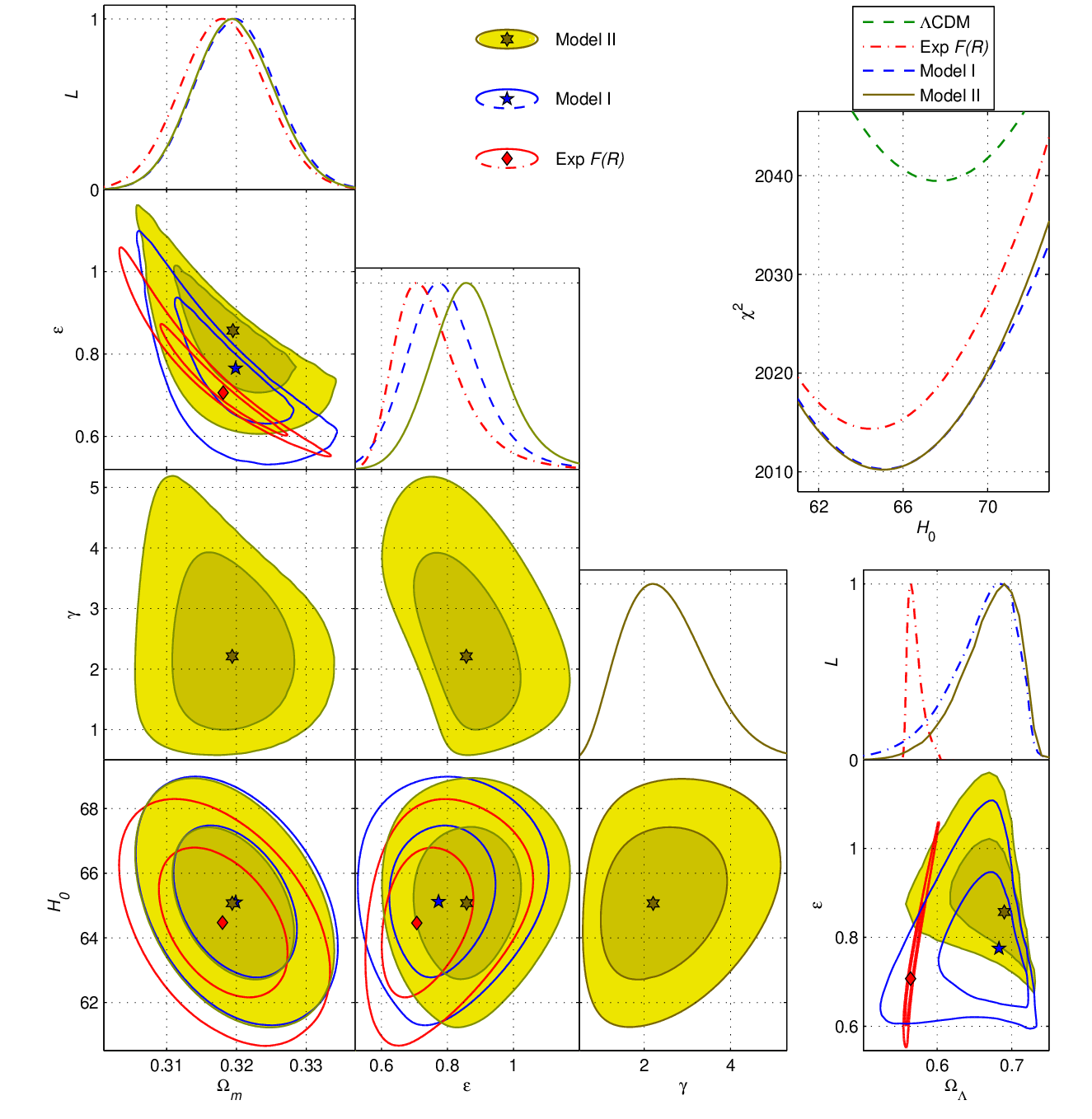}}
\caption{Model II (\ref{ModFR2}) compared with Model I (\ref{ModFR11}) and the
exponential $F(R)$ model (\ref{FRexpon}) in $1\sigma$, $2\sigma$ contour plots of
$\chi^2$, likelihoods $ {\cal L}(\theta_i)$ and $\chi^2(H_0)$ for SNe Ia, CC, CMB and
BAO DESI DR2 data. }
  \label{F4}
\end{figure}

The results of calculations for  Model II  (\ref{ModFR2}) are shown in Table~\ref{Estim}
and in Fig.~\ref{F4} in comparison with Model I and the exponential $F(R)$ model
(\ref{FRexpon}). We see that Model II is a bit more successful than Model I in
$\min\chi^2$, AIC and BIC. The best fits of free parameters for Model II, the
corresponding $1\sigma$, $2\sigma$ contour plots of $\chi^2$ in panels of Fig.~\ref{F4}
are rather similar to their analogs in Model I. In particular, Models I and  II provide
close estimates for $\Omega_m$ and $H_0$, but the best fit
$\varepsilon=0.857^{+0.103}_{-0.103}$ of Model II is larger than for Model I.

Note that both Model I (\ref{ModFR11}) and  Model II
(\ref{fr2212}) tend to $\Lambda$CDM model not only in the limit
$R\to 0$, but also in the limit $\varepsilon\to\infty$. However
for both models the best fits with $1\sigma$ error boxes this
parameter is limited:  $\varepsilon<1$. One may conclude that both
Model I  and  Model II achieve the best $\chi^2$ values when they
are far from their  $\Lambda$CDM limits. We can add that in the
limit  $\varepsilon\to\infty$ or $R\to\infty$, when
$e^{-\varepsilon{\cal R}}\ll 1$ and Model I  and  Model II
converge, their parameters $\alpha$ and $\gamma$ become related:
$\alpha\simeq\gamma^{-1}$. However this relation does not take
place for the best fitted values of model parameters when
$\varepsilon<1$, that can be seen if we compare the estimates from
Table~\ref{Estim}: $\alpha=0.774^{+0.455}_{-0.243}$ and
$\gamma=2.21^{+1.109}_{-0.886}$. The best fits for $H_0$ in
Table~\ref{Estim} for the $F(R)$ models are low, but they confirm
the close estimations for $H_0$ obtained for  other models in
Refs.~\cite{OdintsovSGS:2024,Odintsov:2025kyw}.

The last considered here $F(R)$
model with the  $\Lambda$CDM limit at high $R$ is Model III
(\ref{fr22124}) with the power-law dependence on $R$ in this
limit. Using the notation (\ref{ER}) ${\cal R}=\frac R{2\Lambda}$
we can rewrite its Lagrangian in the case $\beta=2\gamma$ with the
pure  $\Lambda$CDM limit as follows:
\begin{equation}\label{ModFR3}
   F(R)=R+\frac{R^2}{M^2}-\frac{2\Lambda} {1 +\alpha {\cal R}^{-n}}\,.
\end{equation}
This model has two additional free parameters $n$ and $\alpha=
\frac\delta\gamma\big(\frac{m_s^2}{2\Lambda}\big)^n$ leading to
$N_p=5$ (similarly to Model I with $\varepsilon$  and  $\alpha$).
Model III can be also described by the system including
Eq.~(\ref{eqH2}) and the equation,
\begin{equation}\label{eqMod3}
\frac{d{\cal R}}{d\log a}=\alpha_R\frac{\big[\Omega_m^{*}(a^{-3}+ X_r a^{-4})\alpha_R^2
+\Omega_\Lambda^{*}\big(1+\alpha(1-n){\cal R}^{-n}\big)\big]\big/E^2- \alpha_R^2+n\alpha
{\cal R}^{-n}}
 {n\alpha\big[(n+1)\,{\cal R}^{-n-2}+\alpha(1-n)\,{\cal R}^{-2n-2}\big]}\,,
\end{equation}
where $\alpha_R=1+\alpha {\cal R}^{-n}$. Numerical calculations of this system starts
from the $\Lambda$CDM asymptotical initial conditions (\ref{asymLCDM}) at the initial
point
$$ a_\mathrm{ini}=\left[\frac{2\Omega_\Lambda^{*}}{\Omega_m^{*}}\Big({\tilde\delta}^{-\frac1{n+2}}-2\Big)\right]^{-1/3},$$
 it follows from the equality the denominator in Eq.~(\ref{eqMod3}) to a small value $\tilde\delta\sim10^{-9}$.
For all Model I, II and III the values of $\delta=e^{-\varepsilon{\cal R}_\mathrm{ini}}$
or $\tilde\delta={\cal R}_\mathrm{ini}^{-n}$ and $a_{ini}$ were tuned down to the limit
when the resulting expression $H(z)$ appeared to be insensitive to a choice of $\delta$.
For all models the suitable  $\delta$  of order $10^{-9}$ supplies independence of
$H(z)$ on this choice.

These calculations show that viable solutions for Model III appear
to be very close to $\Lambda$CDM solutions during the whole
evolution. Hence, the $\chi^2$ function (\ref{chitot}) calculated
for this model with DR2 DESI BAO data, for a wide range of values
$n$ and $\alpha$ behaves like the  $\Lambda$CDM $\chi^2$. In
particular,
 the two-parameter distribution
$\chi^2(n,\alpha)=\min\limits_{\Omega_m,\Omega_\Lambda,H_0}\chi^2(n,\dots,H_0)$ for
Model III shown in Fig.~5 is practically equal to the constant $\chi^2\sim 2039.4$ at
$\alpha<10$. This constant  $\min\chi^2$ for the $\Lambda$CDM model. The minimum of
$\chi^2$ for Model III is achieved only for very large values of $\alpha$, this result
$\min\chi^2\simeq 2031.11$ is better than the $\Lambda$CDM minimum, this advantage is
kept also for AIC, but Model III strongly concede, if we compare it with Models I and
II. Thus we may conclude that  Model III is unsuccessful in the considered observational
test in comparison with with Models I, II and the exponential model.

 \begin{figure}[th]
   \centerline{ \includegraphics[scale=0.68,trim=5mm 0mm 2mm -1mm]{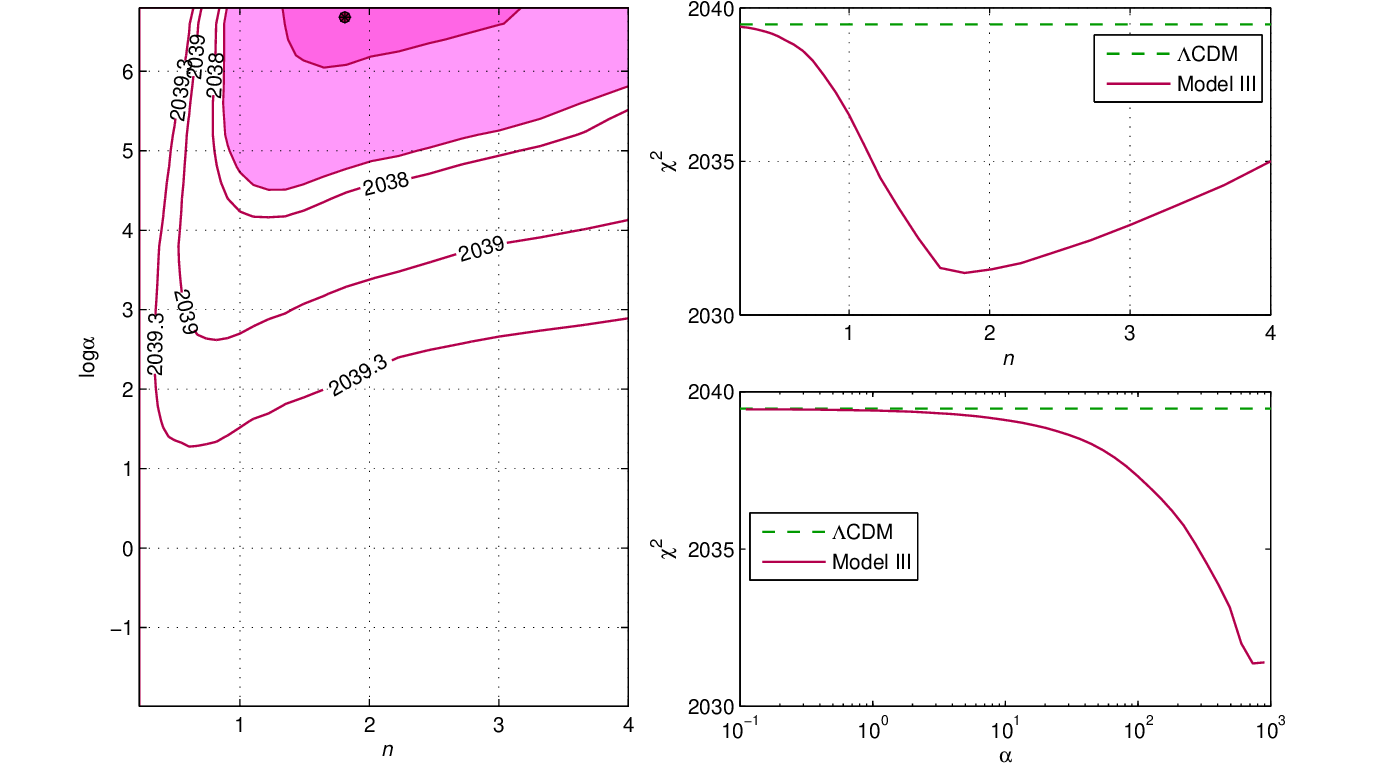}}
\caption{Model III (\ref{fr22124}): $1\sigma$, $2\sigma$ CL
contour plots and contours $\chi^2={}$const in the $n-\log\alpha$
plane and one-parameter distributions $\chi^2(n)$,
$\chi^2(\alpha)$ compared with and $\Lambda$CDM constant minimal
value $\chi^2\sim 2039.46$. }
  \label{F5}
\end{figure}
Model III (\ref{ModFR3}) encounters another problem if we consider the conditions
  $F_R>0$ and $F_{RR}>0$,  which are necessary in order to avoid
anti-gravity effects and supply stability during the matter dominated era
\cite{ElizaldeNOSZ:2011,Chen:2019}. These expressions for Models I, II, III take the
form
 \begin{eqnarray}
 \mbox{Model I}:&& F_R=1+\frac{2R}{M^2}-\frac{\alpha\varepsilon}2 e^{-\varepsilon{\cal R}},\qquad
 F_{RR}=\frac{2}{M^2}+\frac{\alpha\varepsilon^2}{4\Lambda} e^{-\varepsilon{\cal R}}\; \nonumber\\
 \mbox{Model II}:&&F_R=1+\frac{2R}{M^2}-\frac{\gamma\varepsilon e^{-\varepsilon{\cal R}}}{(\gamma+e^{-\varepsilon{\cal R}})^2},\quad
 F_{RR}=\frac{2}{M^2}+\frac{\gamma\varepsilon^2 e^{-\varepsilon{\cal R}}(\gamma-e^{-\varepsilon{\cal R}})}{2\Lambda(\gamma+e^{-\varepsilon{\cal R}})^3}.
 \label{FRR2}\\
 \mbox{Model III}:&&F_R=1+\frac{2R}{M^2}-\frac{\alpha n{\cal R}^{-n-1}}{(1+\alpha{\cal R}^{-n})^2},\quad
 F_{RR}=\frac{2}{M^2}+\frac{\alpha n}{2\Lambda}\frac{(n+1){\cal R}^{-n-2}+\alpha(1-n){\cal R}^{-2n-2}}{(1+\alpha{\cal R}^{-n})^3}.
 \nonumber
 \end{eqnarray}

 \begin{figure}[ht]
   \centerline{ \includegraphics[scale=0.64,trim=5mm 0mm 2mm -1mm]{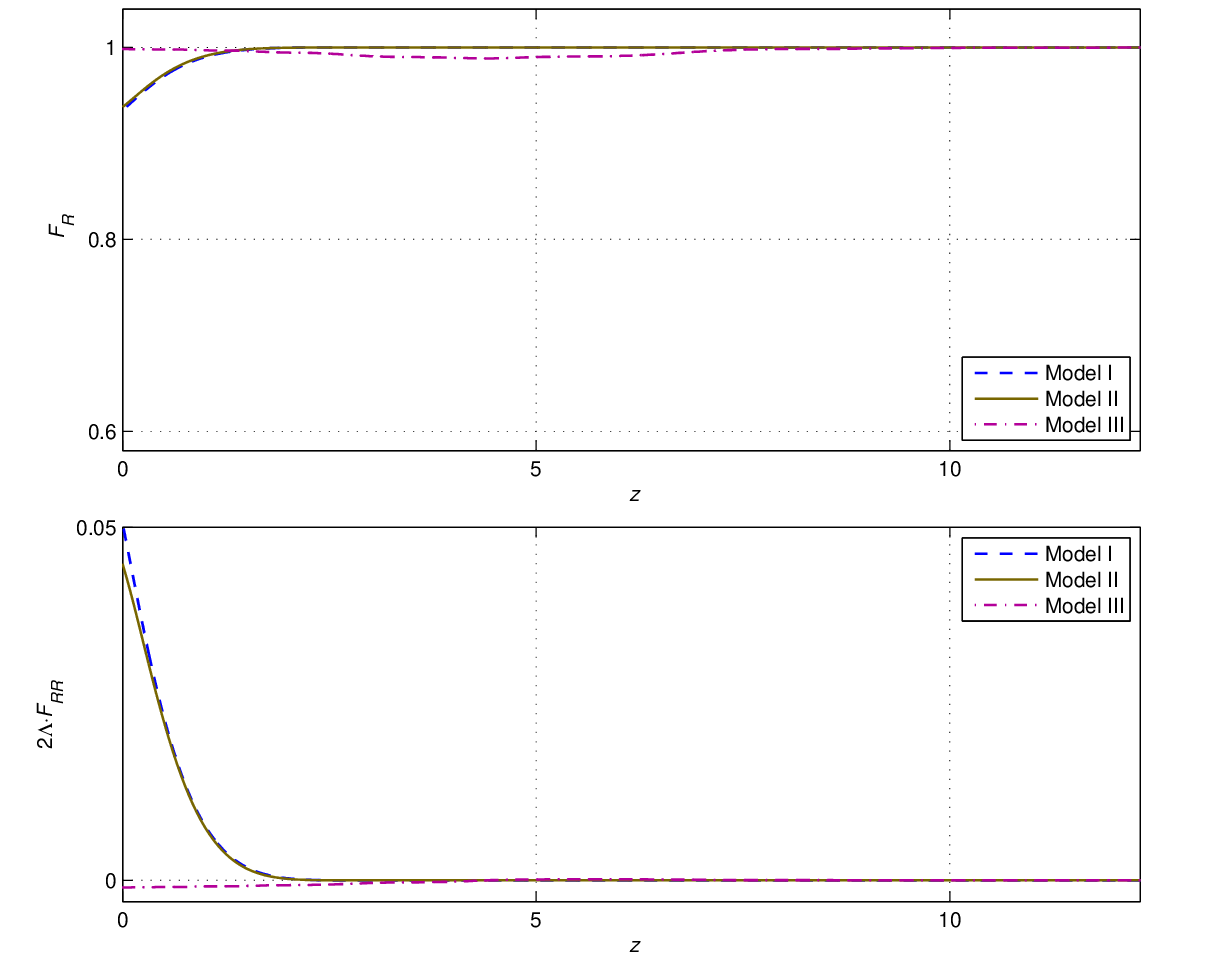}}
\caption{Plots $F_R(z)$ and $2\Lambda\cdot F_{RR}(z)$ for Models I, II, III with  the
best-fit parameters.}
  \label{FFR}
\end{figure}

In Fig.~\ref{FFR} behavior of $F_R(z)$ and $F_{RR}(z)$ are shown for the best-fit
parameters of these models. We see that for all 3 models the condition   $F_R>0$ (and,
moreover,   $F_R>0.9$) is fulfilled, at high $z$ the value $F_R$ is close to 1 and at
very large  $z$ it grows as $1+2R/M^2$. The second derivative $F_{RR}$ is obviously
small and positive for Model I and for Model II (if the best fitted $\gamma=2.21$).
However, for Model III the condition $F_{RR}>0$ can be violated at small $z$ if $n>1$
and $\alpha$ is high enough. This behavior adds arguments against Model III in
comparison with Model I and II.

Note that Models I and II with the best fitted parameters from Table~\ref{Estim} mimic
evolving dark energy with transition from a phantom to a quintessence EoS. As mentioned
above, such a behavior is preferable for describing the DESI DR2 data
\cite{DESI:2025zgx}.

\begin{figure}[ht]
   \centerline{ \includegraphics[scale=0.64,trim=5mm 0mm 2mm -1mm]{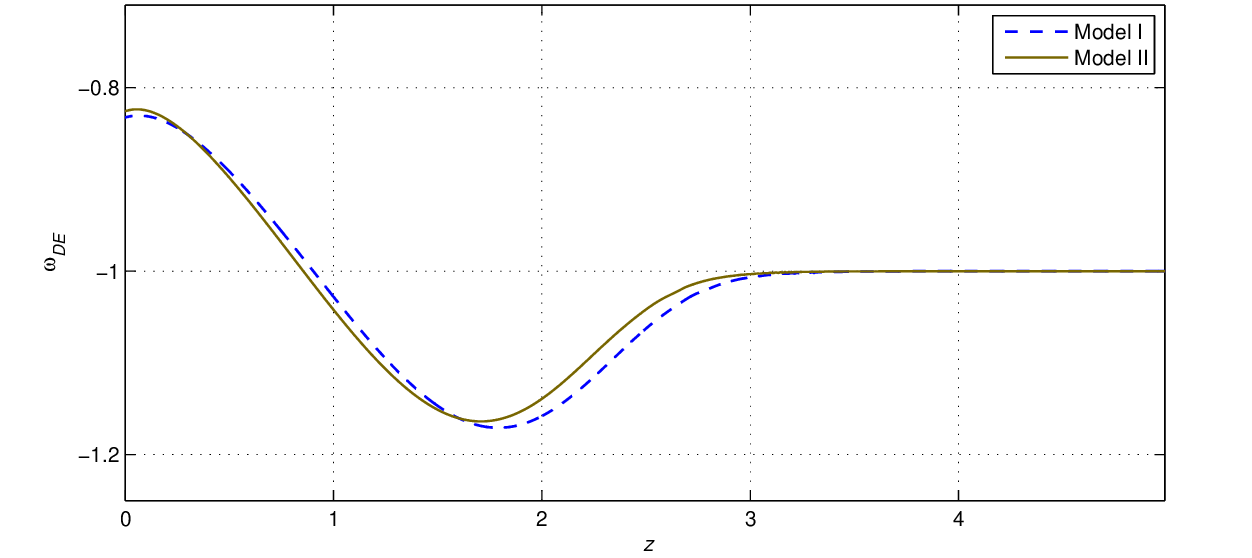}}
\caption{Plots $F_R(z)$ and $2\Lambda\cdot F_{RR}(z)$ for Models I, II, III with  the
best-fit parameters.}
  \label{FwDE}
\end{figure}

Fig.~\ref{FwDE} illustrate evolution of the dark energy EoS parameter (\ref{EoSDE}) az
the function of redshift. For both Models I and II $\omega_{DE}(z)$ tends to $-1$ at
high redshifts in their $\Lambda$CDM asymptotic limit. The effective dark energy evolves
from the  phantom stage ($\omega_{DE}<-1$) and its  transition to the quintessence epoch
with crossing the $\omega_{DE}=-1$ level takes place at $z\simeq0.9$ for Model I and at
$z\simeq0.85$ for Model II.

\section{Conclusions}

In this work we analyzed several viable $F(R)$ gravity models
which provide a unified description of early and late-time
acceleration. These models have a phenomenologically remarkable
behavior since the de Sitter scalaron mass in the Einstein frame
has a monotonically increasing behavior as a function of the
curvature, which renders the models capable of unifying inflation
with the dark energy era. We confronted three such models with the
latest observational data, and we dubbed the models as Model I,
Model II and Model III. We also compared these models with the
$\Lambda$CDM and with a known exponential $F(R)$ gravity model. As
we showed, Model I is advantageous over the exponential model, not
only based on statistics in terms of $\min\chi^2$ but also in AIC.
For Model I we also demonstrated  in Fig.~\ref{F2} the different
behavior of $1\sigma$ and $2\sigma$ CL domains for DR1 and DR2
variants of DESI BAO data: DR1 permits the sets of model
parameters with $\varepsilon<0.5$ and $\alpha>2.5$, but the DR2
excludes these values from the mentioned domains. Regarding the
Model II, it is a bit more successful than Model I from
$\min\chi^2$, AIC and BIC perspective. The best fits of free
parameters for Model II, the corresponding $1\sigma$, $2\sigma$
contour plots of $\chi^2$ where presented in Fig.~\ref{F4} and in
essence, this model is similar to. In particular, both Models I
and II provide close estimates for the $\Omega_m$ and $H_0$, but
the best fit $\varepsilon=0.857^{+0.103}_{-0.103}$ of Model II is
larger than for Model I. Notably, both Model I (\ref{ModFR11}) and
Model II (\ref{fr2212}) are late-time $\Lambda$CDM emulators, not
only in the limit $R\to 0$, but also in the limit
$\varepsilon\to\infty$. However for both these models the best
fits with $1\sigma$ error boxes this parameter is limited:
$\varepsilon<1$. One may thus conclude that both Model I  and
Model II achieve the best $\chi^2$ values when they are far from
their $\Lambda$CDM limits. This is quite important for
phenomenological model building reasons. Regarding the Model III
our analysis indicated that viable solutions for Model III appear
to be very close to $\Lambda$CDM solutions during the entire
late-time evolution. Hence, the $\chi^2$ function calculated for
this model with the DR2 DESI BAO data taken into account, for a
wide range of values of the free parameters $n$ and $\alpha$,
behaves like the $\Lambda$CDM $\chi^2$.  In addition, for Model
III the stability condition $F_{RR}>0$ is violated at small $z$
for the best fitted values $n$ and $\alpha$.

Thus we may conclude that Model III is unsuccessful phenomenologically when it is
confronted with the observational data available compared with Models I, II and the
exponential model.

\section*{Acknowledgments}

This work was partially supported by the program Unidad de
Excelencia Maria de Maeztu CEX2020-001058-M, Spain (S.D.
Odintsov). This research has been funded by the Committee of
Science of the Ministry of Education and Science of the Republic
of Kazakhstan (Grant No. AP26194585) (S.D. Odintsov and V.K.
Oikonomou).

\end{document}